\documentclass[twocolumn,tighten]{aastex631}

\usepackage{amsmath}
\usepackage{amssymb}
\usepackage{ragged2e}
\usepackage{booktabs}
\usepackage{graphicx}

\shorttitle{Chemical diversity in protoplanetary discs and its impact on the formation history of giant planets}
\shortauthors{Pacetti et al.}
\graphicspath{{./}{figures/}}
\begin{document}

\title{Chemical diversity in protoplanetary discs and its impact \\on the formation history of giant planets}

\correspondingauthor{Elenia Pacetti}
\email{elenia.pacetti@inaf.it}

\author[0000-0003-1096-7656]{Elenia Pacetti}
\affiliation{INAF - Istituto di Astrofisica e Planetologia Spaziali (INAF-IAPS), Via Fosso del Cavaliere 100, I-00133, Rome, Italy}
\affiliation{Dipartimento di Fisica, Sapienza Università di Roma, Piazzale Aldo Moro 2, I-00185, Rome, Italy}

\author[0000-0002-1923-7740]{Diego Turrini}
\affiliation{INAF - Istituto di Astrofisica e Planetologia Spaziali (INAF-IAPS), Via Fosso del Cavaliere 100, I-00133, Rome, Italy}
\affiliation{INAF - Osservatorio Astrofisico di Torino, Via Osservatorio 20, I-10025, Pino Torinese (TO), Italy}

\author[0000-0003-1560-3958]{Eugenio Schisano}
\affiliation{INAF - Istituto di Astrofisica e Planetologia Spaziali (INAF-IAPS), Via Fosso del Cavaliere 100, I-00133, Rome, Italy}

\author[0000-0002-9826-7525]{Sergio Molinari}
\affiliation{INAF - Istituto di Astrofisica e Planetologia Spaziali (INAF-IAPS), Via Fosso del Cavaliere 100, I-00133, Rome, Italy}

\author[0000-0002-3911-7340]{Sergio Fonte}
\affiliation{INAF - Istituto di Astrofisica e Planetologia Spaziali (INAF-IAPS), Via Fosso del Cavaliere 100, I-00133, Rome, Italy}

\author[0000-0002-9793-9780]{Romolo Politi}
\affiliation{INAF - Istituto di Astrofisica e Planetologia Spaziali (INAF-IAPS), Via Fosso del Cavaliere 100, I-00133, Rome, Italy}

\author[0000-0002-0472-7202]{Patrick Hennebelle}
\affiliation{AIM, CEA, CNRS, Université Paris-Saclay, Université Paris Diderot, Sorbonne Paris Cité, F-91191, Gif-sur-Yvette, France}

\author[0000-0002-0560-3172]{Ralf Klessen}
\affiliation{Universit\"{a}t Heidelberg, Zentrum f\"{u}r Astronomie, Institut f\"{u}r Theoretische Astrophysik, Albert-Ueberle-Str. 2, D-69120, Heidelberg, Germany}
\affiliation{Universit\"{a}t Heidelberg, Interdisziplin\"{a}res Zentrum f\"{u}r Wissenschaftliches Rechnen, INF 205, D-69120, Heidelberg, Germany}

\author[0000-0003-1859-3070]{Leonardo Testi}
\affiliation{AIM, CEA, CNRS, Université Paris-Saclay, Université Paris Diderot, Sorbonne Paris Cité, F-91191, Gif-sur-Yvette, France}
\affiliation{ESO - European Southern Observatory, Karl-Schwarzschild-Strasse 2, D-85748, Garching bei M\"{u}nchen, Germany}

\author[0000-0001-8060-1890]{Ugo Lebreuilly}
\affiliation{AIM, CEA, CNRS, Université Paris-Saclay, Université Paris Diderot, Sorbonne Paris Cité, F-91191, Gif-sur-Yvette, France}

\begin{abstract}

Giant planets can interact with multiple and chemically diverse environments in protoplanetary discs while they form and migrate to their final orbits. The way this interaction affects the accretion of gas and solids shapes the chemical composition of the planets and of their atmospheres. Here we investigate the effects of different chemical structures of the host protoplanetary disc on the planetary composition. We consider both scenarios of molecular (inheritance from the pre-stellar cloud) and atomic (complete chemical reset) initial abundances in the disc. We focus on four elemental tracers of different volatility: C, O, N, and S. We explore the entire extension of possible formation regions suggested by observations by coupling the disc chemical scenarios with $N$-body simulations of forming and migrating giant planets. The planet formation process produces giant planets with chemical compositions significantly deviating from that of the host disc. We find that the C/N, N/O, and S/N ratios follow monotonic trends with the extent of migration. The C/O ratio shows a more complex behaviour, dependent on the planet accretion history and on the chemical structure of the formation environment. The comparison between S/N* and C/N* (where * indicates normalisation to the stellar value), constrains the relative contribution of gas and solids to the total metallicity. Giant planets whose metallicity is dominated by the contribution of the gas are characterised by N/O*$>$C/O*$>$C/N* and allow for constraining the disc chemical scenario. When the planetary metallicity is instead dominated by the contribution of the solids we find that C/N*$>$C/O*$>$N/O*.	

\end{abstract}

%\keywords{Planet formation -- Protoplanetary discs --  Metallicity -- Chemical abundances -- Abundance ratios -- Extrasolar gaseous giant planets -- Astrochemistry}

%%% ######################################################################## %%%
\section{Introduction} \label{sec:intro}

The past decade has witnessed a huge growth in our knowledge and comprehension of exoplanetary systems, paving the way for a more systematic study of the initial stages of their evolution. Such growth has been achieved thanks to the improved resolution of modern observational facilities. For instance, observations performed with ALMA allowed for the first direct detection of gaps and rings in the gas and dust of protoplanetary discs, which are thought to be the signature of forming giant planets (e.g., \citealp{alma2015, isella2016, fedele2017, fedele2018, andrews2018} and references therein; \citealp{long2018, pinte2018, currie2022}). Improvements in the characterisation of the chemical structure of discs have been made as well, allowing for the first direct comparison between the volatile inventory of extrasolar systems and the Solar System records \citep{drozdovskaya2019, bianchi2019, oberg2021}. Such new evidence combines with the information provided by population studies of exoplanets aimed at investigating the architectures and characteristics of the more than 5000 exoplanets identified to date. The overall emerging picture is that the characteristics of planets (their occurrence, formation pathway, orbital architecture, final mass, and composition) are extremely diverse and uniquely shaped by the physical and chemical properties of the environment in which they formed (e.g., \citealp{madhusudhan2019, zhu2021} and references therein). 

Among the wide variety of worlds discovered so far, giant planets attract considerable interest due to the role they play in shaping the architectures of planetary systems and for their influence on terrestrial planet formation \citep{raymond2014, sotiriadis2018, drazkowska2022}. Our understanding of such planets will soon take a big step forward, thanks to the observations by next-generation telescopes such as JWST and ELT. Moreover, giant planets will be the main target of the upcoming Ariel space mission \citep{tinetti2018, turrini2018, edwards2019b}.

Comparative studies between giant planets in exoplanetary systems and in the Solar System revealed an unexpected diversity in their orbital architectures. Specifically, giant planets are found to cover a wide range of orbital radii, between 0.01 and 100 au\footnote{source: \emph{``The Extrasolar Planets Encyclopaedia"}\\ \url{http://www.exoplanet.eu}}. Although the detection of giant planets far from the star does not raise particular concerns, their existence within 0.1 au poses a great challenge to understanding their formation history. In particular, neither the core accretion \citep{pollack1996}, nor the gravitational instability \citep{boss2000} models, allow a giant planet to form in situ and very close to the star. Therefore, such planets must have migrated inward from their original formation location to the current one. 

What clearly emerges from this result is that the final orbit is simply the end point of the formation process of a giant planet. As such, its properties do not put any constraint on the planet's birthplace and migration history. Instead, \citet{oberg2011} suggested that this information is locked into the final composition of the planet's atmosphere. Such an idea is motivated by the fact that giant planets are expected to form by accreting gas and planetesimals from the surrounding protoplanetary disc. In this regard, migration allows planets to visit regions of the disc with different chemical compositions, as set by the thermal structure of the disc itself. As the chemical abundances in the planet's atmosphere are expected to reflect the composition of the accreted material, they can effectively be used as a proxy for the formation and migration pathways of the planet. A detailed description of how this can be implemented into models of giant planet formation can be found in \citet{turrini2021}, which we refer to as Paper I hereafter, and is summarised in the Appendix~\ref{app:A}. 

In Paper I we simulated the formation and migration of a giant planet in a protoplanetary disc. By focusing on four elemental tracers in the final atmosphere of the planet, namely Carbon (C), Oxygen (O), Nitrogen (N), and Sulphur (S), we built a model to constrain the extent of the migration and the phase of the accreted material. We employed a stationary protoplanetary disc that inherited the composition from its parent cloud. However, a complete picture of the physical and chemical structure of protoplanetary discs is still elusive. One of the still open questions concerns precisely their initial chemical setup. Specifically, it is unclear whether discs inherit their composition from the pre-stellar phase (i.e., from the molecular cloud out of which the host star formed, as assumed in Paper I), in the so called \textit{inheritance} scenario, or experience a complete chemical reset as a result of ionising irradiation from the protostar, known as \textit{reset} scenario. To date, neither the inheritance, nor the reset scenario can completely explain the Solar System record. In fact, comparative studies of meteorites and comets in the Solar System with protostellar objects in protoplanetary discs provide evidence for both of them (e.g., \citealp{oberg2021} and references therein).

In this work, we tested the diagnostic power of our model against different initial chemical conditions for the protoplanetary disc. To further explore the parameter space, we also examined the implications of different levels of ionising radiation reaching the disc midplane. 

Section~\ref{sec:model} provides a description of our planet formation and disc compositional models , further discussed also in the Appendix~\ref{app:A}. In Section~\ref{sec:results}  we present our results in terms of elemental ratios in the disc and in the planet envelope. Specifically, we discuss how the elemental ratios can be used to characterise giant planets in terms of their accretion history (Section~\ref{sec:res_accretion}), migration scenario (Section~\ref{sec:res_migration}), and chemical structure of the birth environment (Section~\ref{sec:res_chemistry}). Finally, in Section~\ref{sec:concl}, we draw the conclusions of our work and summarise the applications of our results.

%%% ######################################################################## %%%
\section{Numerical and compositional model}\label{sec:model}

Paper I studied the link between giant planet formation and composition by coupling a description of the native circumstellar disc, assuming chemical inheritance by the native cloud, with detailed $N$-body simulations of the formation and migration of a giant planet in a dynamically evolving disc of gas and planetesimals. While migrating through different compositional regions of the disc, the planet accretes gas and planetesimals, whose mixture sets the composition of the giant planet and of its extended atmosphere. In this work, we expand the analysis of Paper~I by coupling the outcome of its original simulations with four compositional models for gas and solids in protoplanetary discs. Details on the adopted disc model and on the physical processes and dynamical effects modelled by the simulations are provided in Sections~\ref{sec:discmodel} and \ref{sec:formmodel}. Further details on the planet formation model can also be found in the Appendix~\ref{app:A}. Section~\ref{sec:iondisc} is dedicated to the modelling of the disc ionisation environment. The four compositional models of the protoplanetary disc are described in Section~\ref{sec:compmodel}.

%%% ######################################################################## %%%
\subsection{Disc model: gas and planetesimals} \label{sec:discmodel}

\begin{table}[]
	\caption{Parameters of the disc model.}
	\label{tab:disc}
	\footnotesize
	\begin{tabular}{lcc}
		\toprule
		%\multicolumn{3}{c}{\textbf{Parameters of the disc model}} \\ \midrule\midrule
		\multicolumn{3}{c}{Disc} \\ \midrule\midrule
		%Parameter &	  & Value			\\ \hline
		Stellar mass &	  & $1M_\odot$	\\
		Inner disc radius &	  & $0.1\,$au			\\
		Outer disc radius &	  & $500\,$au			\\
		Disc mass &	  & $0.053M_\odot$ \\
		Disc temperature &	  & $280\,$K at $1\,$au\\
		\midrule
		\multicolumn{3}{c}{Planetesimals} \\
		\midrule\midrule
		Radius &   & $50\,$Km\\
		Density: rock-dominated &	& $2.4\,\text{g cm}^{-3}$\\
		Density: ice-dominated &	& $1\,\text{g cm}^{-3}$\\
		Distribution &   & $1-150$ au\\
		\bottomrule
	\end{tabular}
\end{table}

The protoplanetary disc considered in Paper~I and in this work is modelled over the observed disc HD163296 \citep{isella2016, turrini2019}. The gas surface density and temperature profiles of HD163296 were rescaled to match the total disc mass and the estimated temperature profile of the Minimum Mass Solar Nebula (MMSN; \citealp{hayashi1981}). The physical parameters of the resulting disc are summarised in Table~\ref{tab:disc}. Our disc model assumes radial profiles of gas surface density and temperature on the midplane that are constant in time and parametrised as:
\begin{gather}
	\Sigma_{gas}(r)=\Sigma_0\left(\dfrac{r}{165\,\text{au}}\right)^{-0.8}\exp\left[-\left(\dfrac{r}{165\,\text{au}}\right)^{1.2}\right]\\
	T(r) = T_0\left(\dfrac{r}{1\,\text{au}}\right)^{-1/2},
\end{gather}
where $\Sigma_0=3.3835\,\text{g/cm}^2$ and $T_0=280\,\text{K}$. 

Regarding the solid component of the disc, we assumed that it is distributed between 1 and 150 au with a surface density profile: 
\begin{equation}\label{eq:solids}
	\Sigma_{solids}(r) = \left\{
	\begin{array}{ll}
		2\,Z_i(r)\,\Sigma_{gas}(r) & \quad x \leq 150\,\rm au \\
		0 & \quad x > 150\,\rm au,
	\end{array}
	\right.
\end{equation}
where $Z_i(r)$ is the mass fraction of condensed material. A factor of 2 was introduced as a concentration factor to account for the inward drift of dust and pebbles as a consequence of their dynamical coupling with the gas (see Paper~I and references therein for further discussion). 

Following the approach of Paper~I, we assumed that the bulk of dust in the disc is rapidly converted into planetesimals over a timescale of 1 Myr, as suggested by comparisons between the masses of exoplanetary systems and of the dust and gas in protoplanetary discs \citep[e.g.,][]{manara2018, mulders2021}. The assumption is also consistent with recent observations of dust temporal evolution in discs (see \citealp{testi2022} and \citealp{bernabo2022} for further discussion) and with meteoritic data from the Solar System \citep[e.g.,][and references therein]{scott2007,lichtenberg2022}.

The conversion of dust into planetesimals reduces the efficiency of the gas-grain chemistry, hence slowing down the chemical evolution of the disc itself. When the disc is severely depleted in dust with respect to the initial stages of its evolution, its chemical composition can be reasonably approximated as fixed. We then assumed that the composition of our disc evolves until 1 Myr and remains fixed thereafter (see Section~\ref{sec:compmodel} for details on the compositional model). The implications of physically and chemically evolving discs will be explored in future works.

\begin{figure*}
	\includegraphics[width = \textwidth]{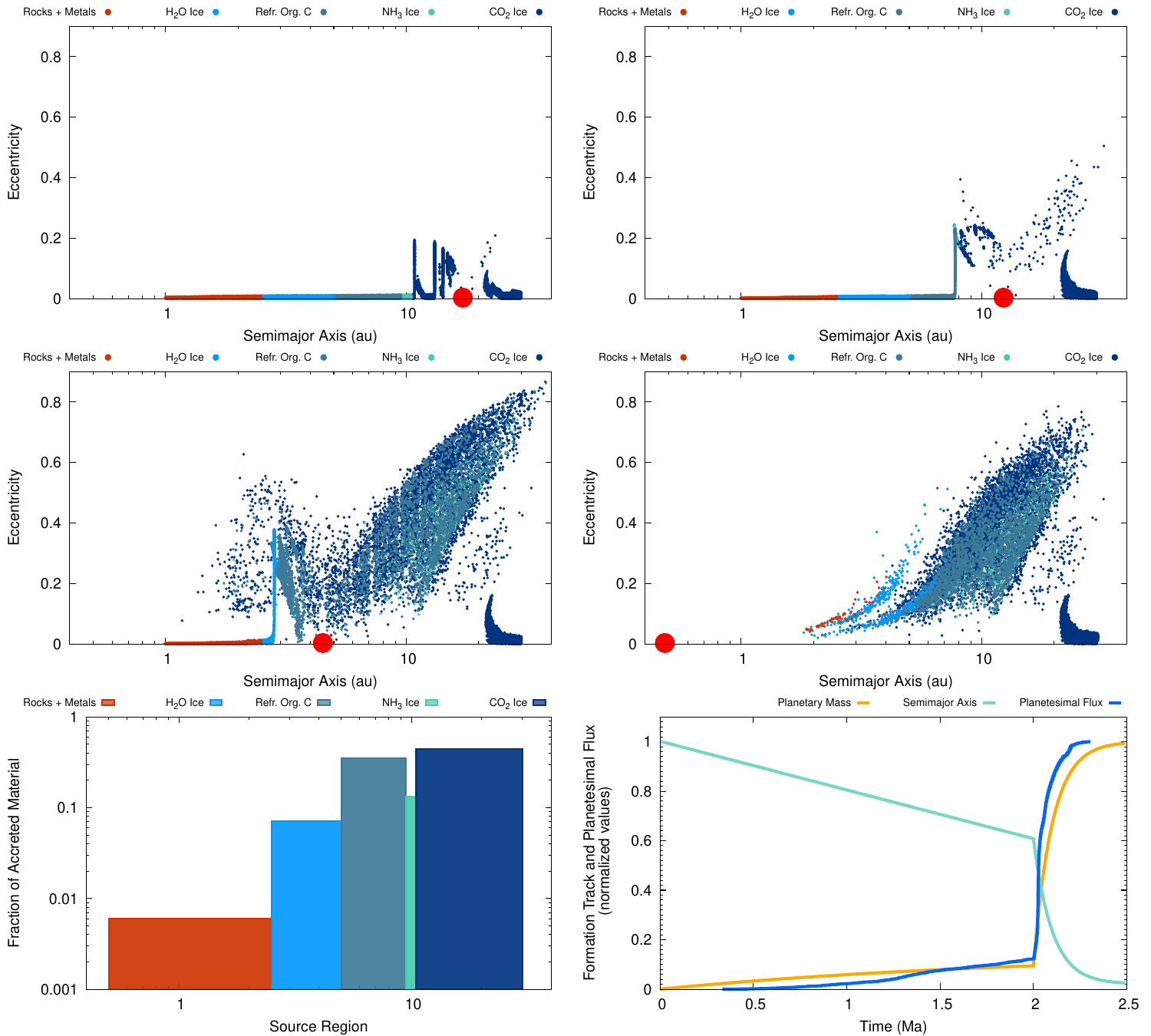}
	\caption{Snapshots of the $N$-body simulations performed with {\sc Mercury-Ar$\chi$es}, describing the formation and migration of a giant planet in a dynamically evolving disc of gas and planetesimals. The four panels at the top and at the centre of the figure show the dynamical evolution of the planetesimals in response to the growth and migration of a giant planet (large red circle) that starts forming at 19~au. From left to right and from top to bottom, the panels show the snapshots of the $N$-body simulations at 0.5, 1.8, 2.1, and 2.5~Myr. Different colours are used to distinguish planetesimals that formed beyond specific snowlines, as indicated in the legend. The two bottom panels are both snapshots taken at 2.5~Myr. The histogram on the left illustrates the fractions of solid material accreted from the different compositional regions of the disc. The plot on the right shows the tracks of the mass growth (orange curve) and the planetesimal accretion (blue curve) normalised to their final values. The green curve follows the evolution of the planet's semimajor axis normalised to its initial value. (An animation of this figure is available.)} 
	\label{fig:mosaic}
\end{figure*}

The planetesimals in our model are all characterised by a fixed radius of $r_p=50$ km \citep{klahr2016, johansen2017} and are divided into two populations, depending on whether they are located within or beyond the water snowline. Planetesimals inside the water snowline are rock dominated and characterised by a density of $\rho_{rock}=2.4$~g~cm$^{-3}$, while those beyond the water snowline are enriched in ices and characterised by a density of $\rho_{ice}=1$~g~cm$^{-3}$. See the Appendix~\ref{app:A} for further discussion.

The radius and density values are used to compute the effects of gas on the planetesimal dynamics. All the planetesimals evolve dynamically under the influence of both the forming giant planet and the disc itself. Specifically, the planetesimals interact gravitationally with the forming giant planet, whose collisional cross section determines whether planetesimals are accreted or scattered by planetary encounters. Moreover, at each location in the disc the planetesimals are subject to two competing forces: the dynamical excitation due to the disc self-gravity and the damping effect of the gas drag. The implementation of the disc self-gravity is based on the analytical treatment for thin discs by \citet{ward1981}, following the approach of \citet{marzari2018} and \citet{nagasawa2019}. The effect of the gas drag on the dynamical evolution of the planetesimals is modelled following the treatment by \citet{brasser2007}, with the updated drag coefficients from \citet{nagasawa2019}. See Paper~I and references therein for more details. 

The planetesimal disc is simulated by means of a set of dynamical tracers distributed randomly between 1 and 150 au with a uniform probability distribution and a spatial density of 2000 tracers/au. Each tracer represents a swarm of planetesimals. The mass of the swarm is computed by dividing the total solid mass in an annular region of the disc (computed by integrating Equation~\ref{eq:solids}) by the number of dynamical tracers it contains. One can then associate the flux of impacting tracers recorded by the simulations with a mass flux of accreted planetesimals on the giant planet (see Paper I for details).

In Figure~\ref{fig:mosaic}, the first four panels from the top are snapshots of the $N$-body simulations showing the dynamical evolution of the planetesimals at different stages of the planet formation and migration history. As the planet forms and migrates, planetesimals from different compositional regions of the disc are dynamically excited and end up being scattered or accreted onto the planet. An example of the normalised flux of accreted planetesimals is shown by the blue curve in the bottom-right panel of Figure~\ref{fig:mosaic}. See Paper~I and the Appendix~\ref{app:A} for more details.

%%% ######################################################################## %%%
\subsection{Formation and migration model} \label{sec:formmodel}

The $N$-body simulations were performed with {\sc Mercury-Ar$\chi$es} \citep{turrini2019, turrini2021}, a high-performance implementation of the hybrid symplectic algorithm of the {\sc Mercury} 6 software from \citet{chambers1999}. Besides the improvements in numerical stability and computational efficiency with respect to {\sc Mercury}, the algorithm of {\sc Mercury-Ar$\chi$es} allows for simulating the mass growth, the radius evolution, and the orbital migration at each stage of giant planet formation, as well as the effects of the disc self-gravity and of the gas drag. We refer the reader to Paper I and the Appendix~\ref{app:A} for more details on the theoretical treatment of these processes and on their numerical implementation in the $N$-body simulations. In the following, we provide a brief overview of what is included in the planet formation and migration model.

We simulated a giant planet that forms and migrates on the disc midplane (i.e., on an orbit with inclination $i=0^\circ$) over a timescale of 3 Myr, growing from a planetary embryo to a Jupiter-like giant planet. We adopted a two-phase approach to model both the mass growth and the evolution of the physical radius of the planet, following the growth tracks from \citet{lissauer2009}, \citet{bitsch2015}, and \citet{dangelo2021} using the parametric approach from \citet{turrini2011, turrini2019}. In the first 2 Myr, the planet accretes its core and extended atmosphere and grows from an initial mass of $M_0=0.1\,M_\oplus$ (mass of the planetary embryo) to a critical mass of $M_c=30\,M_\oplus$, equally shared between the core and the atmosphere. In this phase, the physical radius of the planet grows following the approach described by \citet{fortier2013}, which is based on the hydrodynamical simulations by \citet{lissauer2009}. Over the last 1 Myr, the planet undergoes its runaway gas accretion phase and its mass grows until the final value of $M_F=317.8\,M_\oplus$ (equal to 1 Jovian mass) is reached. At the onset of the runaway gas accretion, the physical radius of the planet starts to shrink due to the gravitational infall of the gas, reaching a final value of $R_I=1.15\times 10^5$ km (equal to 1.6 Jovian radii). An example of the mass growth track normalised to its final value is shown in the bottom-right panel of Figure~\ref{fig:mosaic} (orange curve) for the case of a giant planet that starts forming at 19~au.

To model the migration of the giant planet we adopted the realistic nonisothermal migration tracks from the population synthesis models by \citet{mordasini2015} following a piecewise approach based on the analytical treatments of \citet{han2005} and \citet{walsh2011}. The simulations assume that the protoplanet initially undergoes a damped Type~I migration. During the growth of the planetary core from $M_0$ to the critical value $M_C$, the protoplanet proceeds through a linear regime of slow Type~I migration that accounts for $40\%$ of the total radial displacement of the giant planet. 
When the critical mass is reached and the runaway gas accretion phase begins, the protoplanet enters a faster regime of Type~I migration. Once the protoplanet becomes massive enough to open a gap in the disc, it transitions to the slower Type~II migration regime. These two last phases take the form of a power-law migration regime that accounts for the remaining $60\%$ of the total radial displacement of the giant planet. The green curve in the bottom-right panel of Figure~\ref{fig:mosaic} shows an example of the migration track normalised to its initial value for the case of a giant planet that starts forming at 19~au.

The simulations model a total of six migration scenarios, with the protoplanet starting at 5, 12, 19, 50, 100, and 130~au from the star and ending at 0.4 au. The initial positions span the ranges of the observed architectures of giant planets in the Solar System ($5-10$~au), in exoplanetary systems (from fractions of au to 20~au), and in circumstellar discs ($100-150$~au). The choice of a wide range of distances is meant to investigate the compositional implications of such diverse formation regions.

%%% ######################################################################## %%%
\subsection{Ionisation environment of the disc} \label{sec:iondisc}

As part of our exploration of the parameter space, we investigated how the final composition of the planet changes as we vary the level of chemical activity in the disc. To this aim, we took advantage of the results by \citet{eistrup2016}, who modelled the ionisation environment of the protoplanetary disc as set by two key sources. Specifically, they considered the ionisation from the decay of short-lived radionuclides (SLRs) and from cosmic rays (CRs). For both disc chemical setups they analysed (inheritance and reset, see Section~\ref{sec:compmodel}), they explored a first case of \textit{low} ionisation level, in which SLRs are the only source of ionisation, and a second case of \textit{high} ionisation level, in which an additional contribution from CRs of external origin is also taken into account. This results in a total of four disc chemical scenarios that we used to derive realistic planetary compositions from the outcomes of the simulations of Paper I. Future works will address the dependency of our results on a more detailed treatment of the interaction between the disc and the sources of ionisation \citep{padovani2016, padovani2018, rodgersLee2020}. 

In the low ionisation scenario, the dominant contribution to ionisation comes from the decay products of $^{26}$Al, $^{36}$Cl, and $^{60}$Fe, which have half-lives $t_{\text{half}}$ of 0.74, 0.30, and 2.6 Myr, respectively \citep{cleeves2014}. \citet{eistrup2016} adopted a simplified version of the analytical prescription given in Equation~30 by \citet{cleeves2013b} for the ionisation rate per H$_2$ molecule at the disc midplane:
\begin{equation}\label{eq:slrrate}
	\zeta_\text{SLR}(r)=(1.25\times 10^{-19}\, \text{s}^{-1})\,\left(\dfrac{\Sigma(r)}{\text{g}\,\text{cm}^{-2}}\right)^{0.27},
\end{equation}
where $\Sigma(r)$ is the surface density of the disc as a function of the radius $r$. Equation~\ref{eq:slrrate} does not account for the time decay of the ionisation rate. Therefore, in the SLRs-dominated environment by \citet{eistrup2016}, the ionisation rate is constant in time and higher in the inner and denser regions of the midplane. 

For the high ionisation scenario, in addition to SLRs, \citet{eistrup2016} included a contribution of cosmic rays originating from outside the system. They parametrically modelled the CR ionisation rate per H$_2$ molecule as: 
\begin{equation}\label{eq:crrate}
	\zeta_\text{CR}(r)=(1\times 10^{-17}\, \text{s}^{-1})\,\cdot\exp\left(\dfrac{-\Sigma(r)}{96\,\text{g}\,\text{cm}^{-2}}\right),
\end{equation}
where the characteristic value of $\zeta_\text{CR}\sim10^{-17}\, \text{s}^{-1}$ is the interstellar rate that is typically assumed in models of disc chemistry \citep{webber1998, cleeves2014}. The addition of CRs allows for higher ionisation in the outer disc, where the surface density is lower and CRs can more efficiently penetrate into the midplane. 

The aim of considering the two scenarios is to investigate the impact of ionisation-driven chemical activity on the chemical composition of the disc midplane. Both the prescriptions given by Equations~\ref{eq:slrrate} and \ref{eq:crrate} should be interpreted as conservative estimates of the rates that regulate the ionisation environment of a typical midplane. A recent study by \citet{padovani2018} suggests that the CR ionisation rate in high-density environments is higher than the typically assumed value of $10^{-17}\, \text{s}^{-1}$. Moreover, the model by \citet{eistrup2016} includes only attenuation due to the disc's surface density. Works by \citet{cleeves2013a, cleeves2013b, cleeves2014} revealed that the CR flux at the disc's surface may be strongly attenuated by winds and/ or magnetic fields. In particular, modulation by stellar winds may reduce the CR flux reaching the midplane by many orders of magnitude ($\zeta_\text{CR}\,\lesssim\,10^{-20}\, \text{s}^{-1}$), leaving SLRs as the dominant midplane ionisation source \citep{cleeves2013a}. Nevertheless, as it was mentioned before, the abundances of SLRs do evolve with time. As pointed out by \citet{cleeves2013b}, the SLRs ionisation rate scales with the inverse of $t_{\text{half}}$ and long-lived radionuclides (e.g., $^{40}$K) only produce ionisation rates of the order of $\sim10^{-22}\, \text{s}^{-1}$ or even less. Moreover, although there is evidence of an enhanced abundance of SLRs in the early Solar System, little is still known about their actual contribution to the total ionisation rate in the midplane of a typical disc \citep{cleeves2013b}.

Regarding other sources of ionisation, \citet{cleeves2013a} showed that photoionisation from stellar and interstellar UV acts largely on C-bearing species in the upper atmosphere of the disc, leaving the midplane essentially unaffected. On the other hand, X-ray photons from the star penetrate deeper and dominate in the intermediate layers of the disc. Only the scattered component of stellar X-rays is able to reach the midplane \citep{ercolano2013}. However, for a stellar X-ray luminosity of $L_X=10^{29.5}\, \text{erg s}^{-1}$, the scattered X-rays are expected to produce ionisation rates falling in the range $\zeta_\text{CR}\sim(1-10)\times10^{-21}\, \text{s}^{-1}$, far below the rates associated with SLRs and CRs, with minimal impact on the midplane chemistry \citep[see][and Figures~2 and 3 therein]{cleeves2014}. Concerning X-ray background fields, as in the case of a protoplanetary disc embedded in a cluster, recent works revealed that these would have relatively little impact on disc chemistry \citep{meijerink2012,rab2018}. In particular, for an ordinary disc in a typical low-mass star-forming region, the ionisation rate produced by an X-ray background flux of $2\times10^{-4}\text{erg cm}^{-2}\,\text{s}^{-1}$ would equal the interstellar $\zeta_\text{CR}$ only in the outer disc, at $r\sim200$ au from the central star \citep[see][and Figure~10 therein]{rab2018}.

%%% ######################################################################## %%%
\subsection{Compositional model of the disc}\label{sec:compmodel} 

\begin{figure*}
	\includegraphics[width = \textwidth]{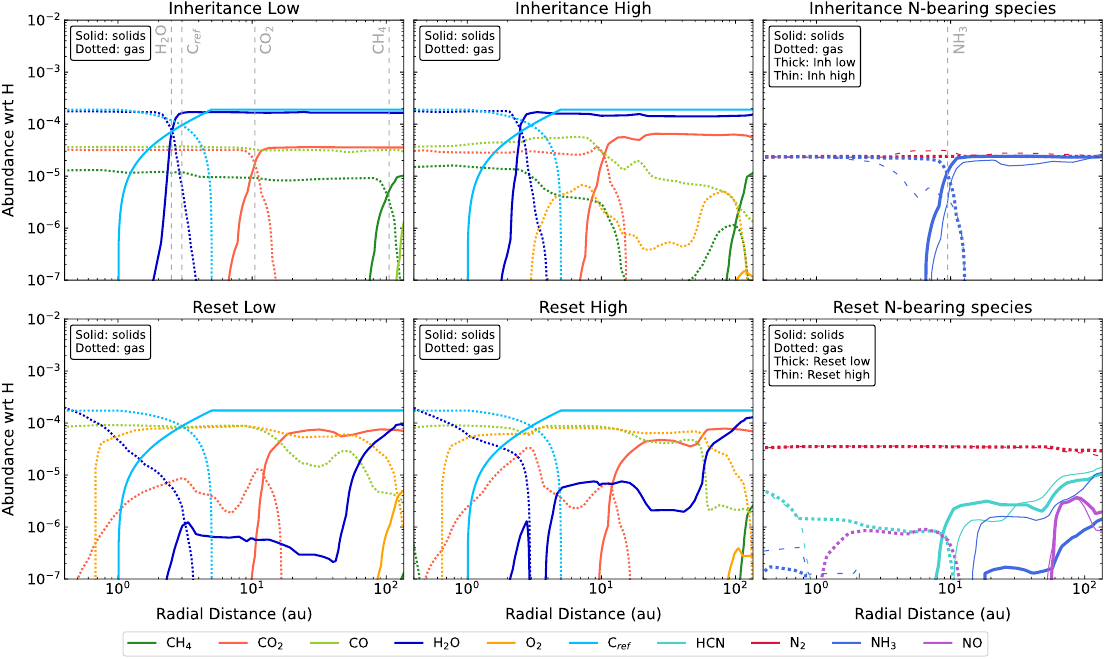}
	\caption{Radial profiles of molecular abundances of the key volatile C-, O-, and N-bearing species with respect to total H atoms. The profiles are taken from \citet{eistrup2016} and properly rescaled to account for the different disc and the different compositional model considered in this work. In all panels, solid and dotted lines indicate the solid and gas phases, respectively. The top three panels describe the two inheritance scenarios, while the bottom three ones describe the two reset scenarios. The left-hand and central panels show the abundances of the key volatile molecules carrying C and O, including the contribution of refractory organic carbon (C$_{ref}$). N-bearing molecules are shown in the two right-hand panels for visual ease. Here, the low (SLRs only) and high (SLRs and CRs) ionisation levels are shown in the same plot by means of thick and thin lines, respectively. The grey, dashed, vertical lines in the plots of the inheritance low scenario indicate the position of the snowlines of H$_2$O (2.5 au), C$_{ref}$ (3 au), NH$_3$ (9.4 au), CO$_2$ (10.5 au), and CH$_4$ (105 au). CO and N$_2$ condense beyond 130 au in our model and their snowlines are therefore not seen. Snowlines are omitted in the other scenarios, as the complexity of the chemistry requires the introduction of multiple snowlines.} 
	\label{fig:profiles}
\end{figure*}

\begin{figure*}
	\includegraphics[width = \textwidth]{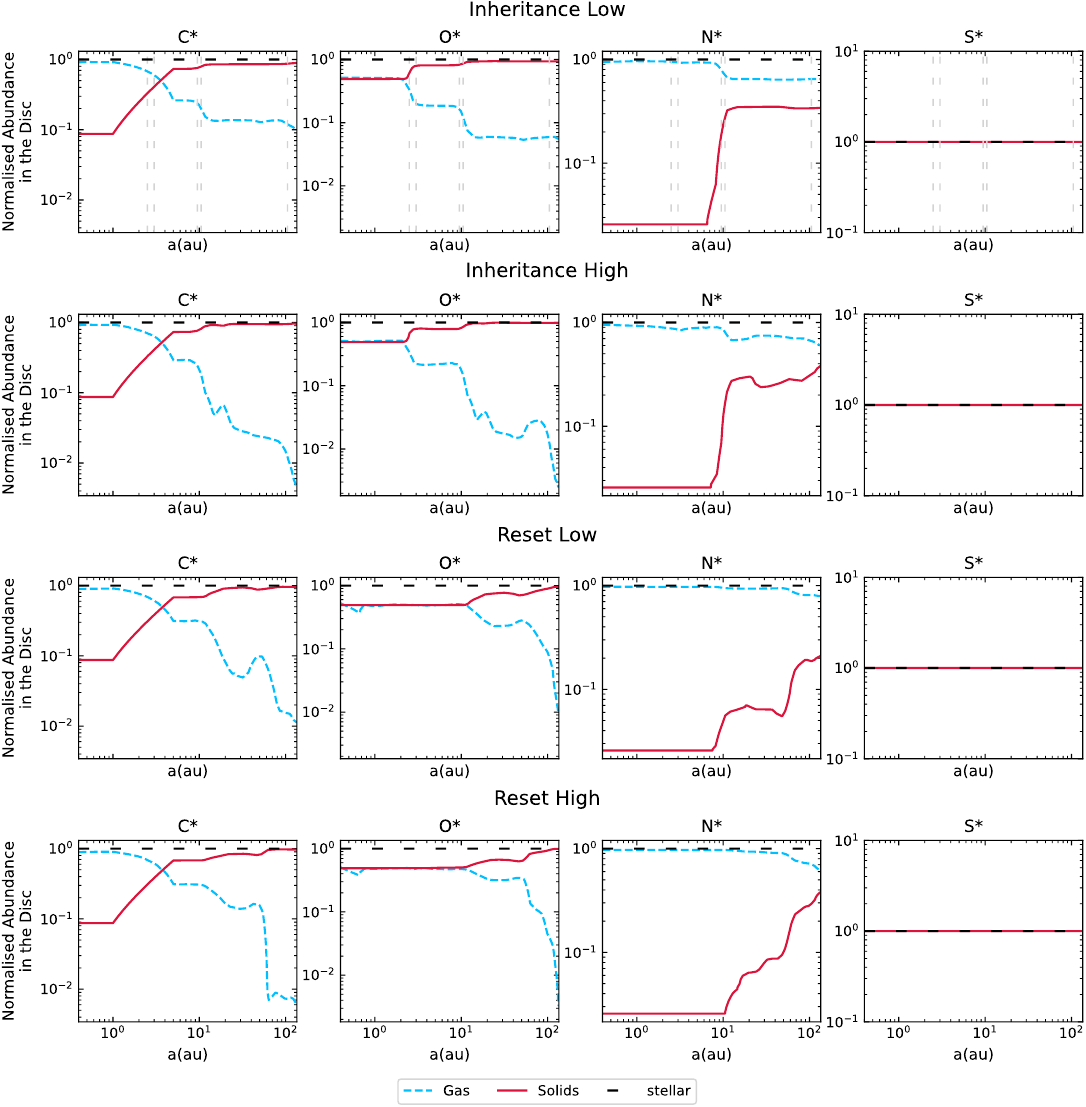}
	\caption{Elemental abundances of C, O, N, and S in the disc, normalised to their respective stellar value, for the gas (dashed lines) and the solid (solid lines) phases (the latter including rocks, organics, and ices). The trends are extracted from the radial profiles of molecular abundances in the four chemical scenarios. The grey, dashed, vertical lines in the top four panels indicate, from left to right, the locations of the snowlines of H$_2$O (2.5 au), refractory organic carbon (3 au), NH$_3$ (9.4 au), CO$_2$ (10.5 au), and CH$_4$ (105 au). Such snowlines are unequivocally defined only in the inheritance scenario with low ionisation level and for that reason they are omitted in the other scenarios. The black, dashed, horizontal line at 1 indicates the reference stellar value in all plots.}
	\label{fig:disc_comp_elem}
\end{figure*}

To characterise the chemical environment in which the giant planet forms and migrates, we used an updated version of the compositional model presented in Paper~I. The model quantifies the composition of the disc over its key components: rocks, organics, ices, and gas. In this work, we introduce a more realistic abundance profile for the refractory organic carbon and we revise the treatment of the volatile component to allow for different evolutionary scenarios and ionisation levels. Here, we outline the key features of the model and we provide details on its refinements. We refer the reader to Paper I for a complete discussion. 

We adopted the protosolar composition for both the star and its hosting protoplanetary disc. The protosolar elemental abundances, which characterise the original mixture of gas, were taken from \citet{asplund2009} and \citet{scott2015a,scott2015b}. Overall, they result in mass fractions $X=0.7148$, $Y=0.2711$, and $Z=0.0141$, of H, He, and heavy elements, respectively. The gas composition sets the initial conditions for the condensation sequence across the disc, which regulates the distribution of the elements among the different phases (gas and solid) and carriers (rocks, organics, gas, and ices). As in Paper I, we focus on four tracing elements: C, O, N, and S. Their partitioning between the three solid components and between the solid and gas phases is derived from Solar System data and recent results from astrochemical models, as discussed below. 

%%% ######################################################################## %%%
\subsubsection{Refractory elements and rocks}\label{sec:rocks}

Based on meteoritic \citep{lodders2010,palme2014} and cometary \citep{leRoy2015, rubin2019, rubin2020, altwegg2019} data, we assumed that rock-forming elements are subtracted from the gas phase and locked into rocks in meteoritic proportion. In this framework, the comparison between meteoritic \citep{lodders2010,palme2014} and protosolar abundances \citep{asplund2009, scott2015a,scott2015b} reveals that rock-forming elements account for a mass fraction $Z_{rock}=6.67\times 10^{-3}$ of the total gas in the disc. By subtracting the meteoritic abundances from the protosolar ones, we computed the initial abundances of the elements that remain in the gas phase as volatiles. Their partition across the different molecular carriers will be discussed in Section~\ref{sec:volat}. 

Focusing on the four elemental tracers considered in this work, the residual gas in our model contains $51\%$ of the protosolar O and the majority of C ($91\%$) and N ($97\%$). The fact that almost half of the total O is trapped in rocks while N remains almost entirely in the gas phase, has important consequences for the final elemental ratios in giant planet atmospheres. One of them is that the N/O ratio of giant planets that derived most of their metallicity from the accretion of gas is always superstellar (see Section~\ref{sec:results} for a more detailed discussion of this and other implications). While the gas remains substantially enriched in C, O, and N, the totality of S is trapped in solids as chondritic rocks across the whole disc. In this work, we assumed rocks to be fully incorporated into planetesimals at the onset of our simulations, which makes S particularly effective in tracing the planetesimal accretion (see Section~\ref{sec:res_migration} for further discussion). 

%%% ######################################################################## %%%
\subsubsection{Volatiles}\label{sec:volat}

The partition of volatiles across the different phases and carriers in the protoplanetary disc is based on the astrochemical models by \citet{eistrup2016}, who simulated four different scenarios obtained by varying the chemical initial conditions in the disc and the incident flux of ionising radiation. Specifically, they explored  two disc chemical scenarios called  \emph{inheritance} and \emph{reset}. In the inheritance scenario, the disc is assumed to have inherited its initial composition from the original molecular cloud. Here, the initial conditions of the chemistry are set by the abundances of H, He, H$_2$, and eight key volatile molecules \citep[see][and Table. 1 therein]{eistrup2016}. In the reset scenario, the disc chemistry is instead  completely reset due to heating from the protostar and the molecular gas of the disc, except for H$_2$, is fully dissociated into atoms. The chemical network is therefore evolved starting from atomic initial abundances of H, He, C, O, N, and S, plus molecules of H$_2$. Each of the two scenarios was analysed under condition of both \emph{low} and \emph{high} levels of ionisation, as described in Section~\ref{sec:iondisc} and in \citet{eistrup2016}.

To model the distribution of C, O, and N across their molecular carriers in the four scenarios discussed above, we employed the radial abundance profiles by \citet{eistrup2016} of the main volatile molecules in the gas and ice phases, produced by their full chemical network after 1~Myr of evolution. In doing so, we had to account for two major differences between our compositional model and the one by \citet{eistrup2016}. First, \citet{eistrup2016} consider a disc with a single component of volatiles, which evolves in the gas phase and partly condenses into ices. As mentioned before in this section, our compositional model describes a three-component disc in which rocks and refractory organic carbon are formed alongside volatiles. Second, \citet{eistrup2016} start from the total initial abundances of volatiles that are modelled over those of the interstellar medium. In our model, the three components originate from a mixture of gas with protosolar elemental abundances. We then performed a rescaling of the radial abundance profiles by \citet{eistrup2016} to account for the additional components in the disc and to ensure that the protosolar abundances are retrieved throughout the disc for each tracing element.

To perform the scaling, we took advantage of the fact that chemical networks characterised by very different timescales can be treated independently of each other. Details of the adopted procedure are provided in the remainder of this section.

For the two inheritance scenarios, we followed the approach described in Paper I. \citet{eistrup2016} report a total initial abundance of N ($6.30\times 10^{-5}$) smaller than that available to form volatiles in our model ($7.22\times 10^{-5}$). To ensure the conservation of the total mass of N, we then scaled up the total abundance of N and the abundance profiles of NH$_3$, and N$_2$ from \citet{eistrup2016} by a factor of 1.15. When it comes to C and O, the two of them share a joint chemical network, which in turn requires using the same scaling factor for C- and O-bearing volatiles in order to preserve their relative proportions in the astrochemical model. We then scaled down the total abundances of both C and O and of their volatile molecular carriers by a factor of 0.53. The factor was computed by comparing the total initial abundance of volatile O ($5.20\times 10^{-4}$) reported by \citet{eistrup2016} with the one available in our model to form volatiles ($2.74\times 10^{-4}$), which takes into account the fact that almost half of the protosolar oxygen is sequestered into rocks (see also Section~\ref{sec:rocks}). The remaining abundance of $1.9\times 10^{-4}$ of C ($\sim 60\%$ of the total protosolar abundance) is associated with refractory organic carbon (see also Paper I for further discussion), consistently with cometary and interstellar medium (ISM) data \citep{bergin2015}.

When considering the reset scenarios, two additional N-bearing species, HCN and NO, reach non-negligible abundances throughout the disc. However, both HCN and NO form more rapidly than NH$_3$ and N$_2$. Specifically, HCN and NO form quickly through gas phase reactions \citep{eistrup2016, schwarz2014}, alongside the other volatile C- and O- bearing molecules (e.g., CO and CO$_2$). On the other hand, the formation of NH$_3$ is governed by the slower gas-grain chemistry. The formation of N$_2$ occurs at an intermediate timescale. In the network of N, is therefore reasonable to assume that the formation of HCN and NO, and of NH$_3$ and N$_2$ proceeds through separate pathways. Specifically, the formation of HCN and NO occurs at early stages and is limited by the availability of atomic C and O in the gas phase. NH$_3$ and N$_2$ form later on from the residual N left by the previous reactions. When calibrating the radial abundance profiles, this translates into using different scaling factors for the formation pathways of HCN and NO, and of NH$_3$ and N$_2$. Specifically, we scaled the abundances of HCN and NO down by the same factor we applied to O- and C-bearing species, this factor being 0.53 as in the case of the inheritance scenarios. The excess of N at each location in the disc was then used to compute the scaling factor of N$_2$ and NH$_3$. As in the inheritance, also in the reset scenarios a fraction of atomic C is associated with refractory organic carbon. In this case, the total abundance is $1.7\times 10^{-4}$. The rescaled molecular abundance profiles used in this work are plotted in Figure~\ref{fig:profiles} for both the inheritance and the reset scenarios.

The most noticeable effect of varying the level of chemical activity is that the snowlines become less unequivocally defined, as multiple snowlines for the same molecule can exist at different locations in the disc. The O$_2$ carries a significant fraction of O across all scenarios, except for the inheritance low scenario, where the abundance of O$_2$ drops below $10^{-7}$. Among the C-bearing species, the dominant contribution comes from organic compounds, whose origin is discussed in more details in Section~\ref{sec:reforg}. The N$_2$ is the main carrier of N in all scenarios. As its snowline is located beyond 130 au for the adopted disc thermal profile, a large fraction of N remains in the gas phase across most of the disc. The fraction of N that condenses as ice is carried by NH$_3$ in the inheritance scenarios and by NH$_3$, HCN and NO in the reset ones. 

The focus of our compositional model is on the distribution of the four tracing elements, C, O, N, and S across the gas and solid phases in the disc, not on the distribution of their respective molecular carriers, as represented by the radial abundance profiles. When it comes to volatiles, the relevant quantity to consider is therefore the abundance of each tracing element, computed as the sum of the abundances of all the molecular carriers of that specific element. The partition of the molecular abundance profiles in Figure~\ref{fig:profiles} among C, O, N, and S, is shown in Figure~\ref{fig:disc_comp_elem} for both the inheritance and the reset scenarios. The dashed lines indicate the contribution from the gas phase, while the solid lines indicate that of solids. Note that the solid phase includes also the contribution from the rocks (see Section~\ref{sec:rocks}), in addition to the ices and the organics. The elemental abundances are normalised to their respective stellar value, indicated by the black, dashed, horizontal line. We adopt the superscript * in the notation as a general convention to distinguish the normalised abundances from the absolute ones. In this scale, S is at the stellar level throughout the disc, being entirely locked into rocks. For the other elements, the trends reflect the phase changes of the disc material across the multiple snowlines. The major differences among the four scenarios are in the gaseous O, which shows higher abundances in the reset scenarios with respect to the inheritance ones. Such differences have a significant impact on the planetary elemental ratios that are computed from O. For instance, for planets that derived their metallicity from the accretion of gas in the reset scenarios, the large fraction of O locked in the gas phase translates into substellar C/O ratios. See Section~\ref{sec:res_accretion} for further discussion of this and other effects.

%%% ######################################################################## %%%
\subsubsection{Refractory Organics}\label{sec:reforg} 

Following \citet{cridland2019} and previous authors \citep[e.g.,][]{thiabaud2014, bergin2015, mordasini2016}, our model includes a  refractory C-bearing component as suggested by the comparison between the carbon deficit observed in the Earth \citep{allegre2001} and in Solar System meteorites  \citep{wasson1988, bergin2015}  with respect to the ISM and comets (\citealp{bergin2015} and references therein; \citealp{bardyn2017}). Such component accounts for $\sim 60\%$ of the ISM carbon, yet it is unclear whether C is in the form of graphite, amorphous carbon grains, or organics. We adopted the terminology by \citet{thiabaud2014} and \citet{isnard2019}, and we refer to this C reservoir as refractory organic carbon.

For the bulk composition of the inner Solar System to be markedly carbon poor, refractory organic carbon needs to be destroyed in the inner Solar Nebula. In particular, following \citet{lee2010}, we assumed that it is completely destroyed and released into the gas phase within 5~au. Following \citet{mordasini2016} and \citet{cridland2019}, we introduced a snowline for refractory organic carbon with 50$\%$ condensation at 3 au, in both the inheritance and reset scenarios. The resulting abundance profile for the condensed phase is zero inside 1~au, then increases  linearly up to 5~au, where it reaches the value of $1.9\times 10^{-4}$ in the inheritance scenario and of $1.7\times 10^{-4}$ in the reset one (see Section \ref{sec:volat}), and remains constant  thereafter. The distribution of refractory organic carbon across the disc is shown Figure~\ref{fig:profiles} by means of light-blue curves. 

%%% ######################################################################## %%%
\section{Results and Discussion}\label{sec:results} 

As introduced in Section~\ref{sec:model}, in order to explore the implications of the disc chemistry for the chemical composition of the planetary atmospheres, we coupled the outcome of the $N$-body simulations from Paper I with the four compositional models of the disc. The simulations describe a planetary embryo that starts its growth and migration in different compositional regions of the disc, specifically migrating from 5, 12, 19, 50, 100, and 130 au to 0.4 au (see also Figure~\ref{fig:mosaic}). Each simulation traces the accreted masses of both gas and planetesimals at each location along the migration pathway.

For each simulation and compositional model we proceeded as follows. We first used the radial abundance profiles of the main molecular carriers of C, O, N, and S (Figure~\ref{fig:profiles}) and the relevant atomic weights to extract the individual mass contributions of the four tracing elements from the accreted masses of gas and solids. Under the assumption of homogeneous mixing in the planetary envelope (see Paper I for discussion), for each element we computed the total accreted masses of gas and solids along the formation pathway. We then converted such masses into total abundances with respect to H atoms. The results were used to compute the elemental ratios C/O, C/N, N/O, and S/N, in the final atmosphere of the planet, for both the gas and solid phases. As discussed in Paper I, when considering both contributions from the accreted gas and solids, the giant planet is \emph{solid enriched}. In this case, the planet envelope is characterised by superstellar abundances of most if not all the elements. On the other hand, when considering only the accretion of gas, the giant planet is \emph{gas dominated}. For the adopted compositional model of the disc, gas-dominated giant planets are characterised by substellar abundances (see Paper I and \citealp{turrini2021b} for further discussion of alternative scenarios).

The final step in our analysis was to compute the atmospheric elemental ratios normalised to their stellar counterparts. In this scale, values of the ratios equal to~1 are reflective of a giant planet whose chemical composition matches that of its host star. In this sense, normalisation provides a tool to measure the deviation of the planet's composition from the stellar one. The results can then be generalised to systems in which the composition of the host star differs from the solar one considered in this work. To differentiate between absolute and normalised ratios, we refer to the latter as C/O*, C/N*, N/O*, and S/N*. 

\begin{figure*}
	\includegraphics[width = \textwidth]{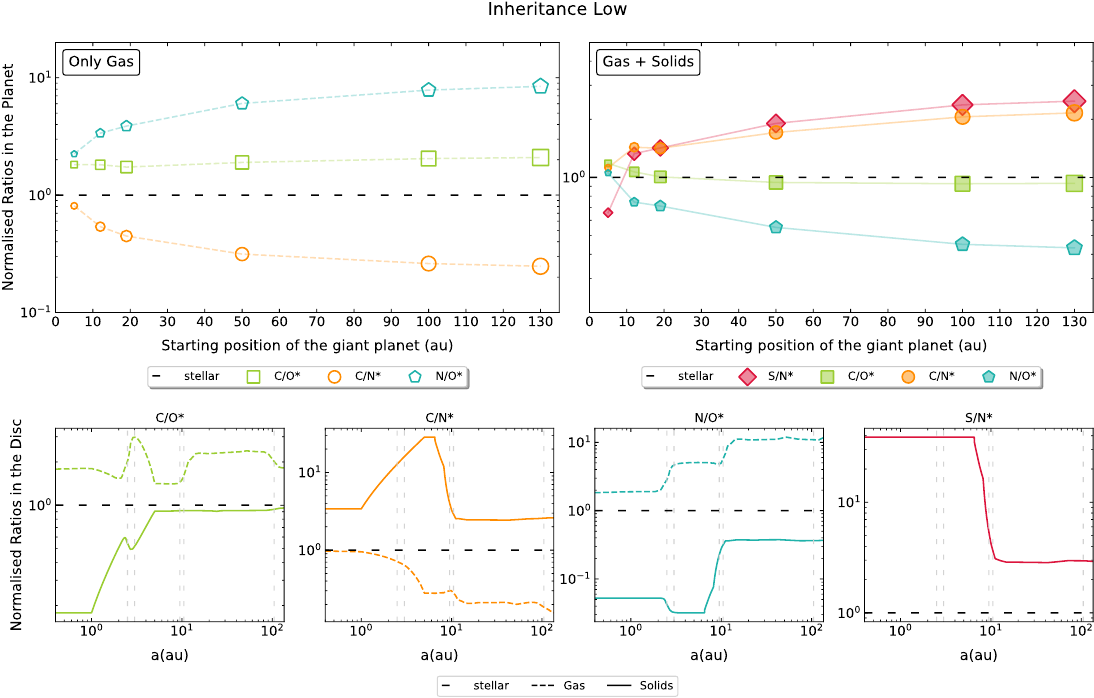}
	\caption{Normalised elemental ratios in the final atmosphere of the planet (top two panels) and in the disc (bottom four panels) in the inheritance scenario with low ionisation. The top two panels show results for the six migration scenarios, corresponding to initial semimajor axes of 5, 12, 19, 50, 100, and 130 au. The increasing size of the markers maps the increasing distance travelled by the planet in the six migration scenarios. Linking lines are shown only for visual ease. The plot on the left (with empty markers) describes a gas-dominated giant planet, while the plot on the right (filled markers) is for solid-enriched ones. The bottom four panels show the trends of the elemental ratios on the disc for the gas (dashed lines) and the solid (solid lines) phases. The grey, dashed, vertical lines, from left to right, indicate the position of the snowlines of H$_2$O (2.5 au), refractory organic carbon (3 au), NH$_3$ (9.4 au), CO$_2$ (10.5 au), and CH$_4$ (105 au). The black, dashed, horizontal line at 1 indicates the reference stellar value in all plots.}
	\label{fig:ternary1}
\end{figure*}

\begin{figure*}
	\includegraphics[width = \textwidth]{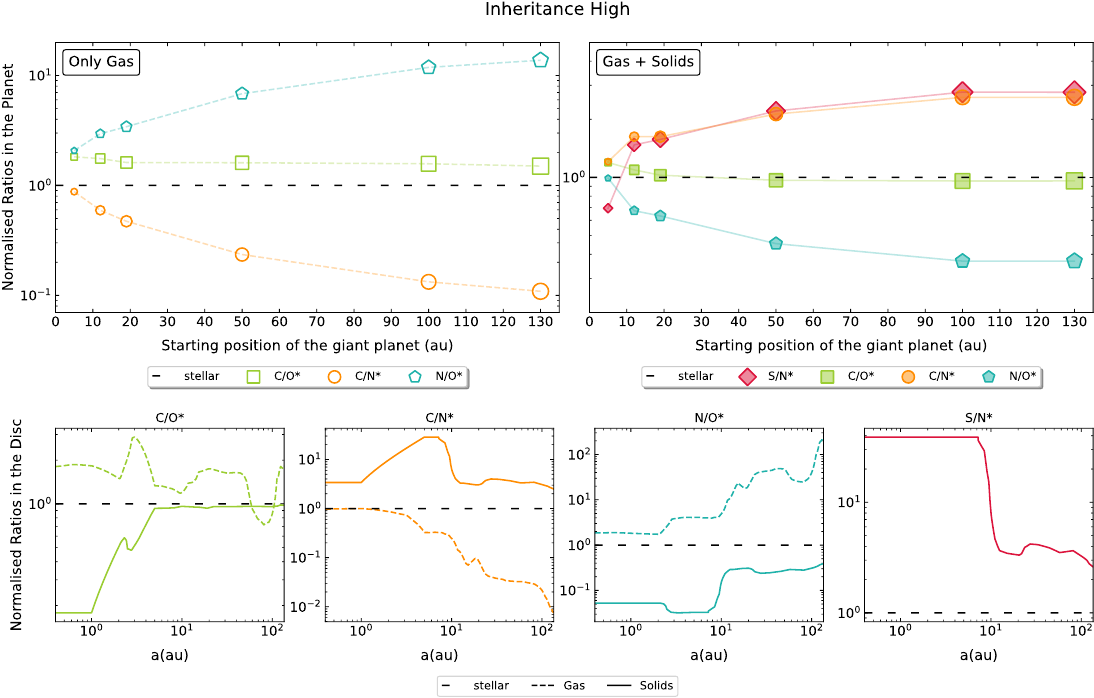}
	\caption{Normalised elemental ratios in the final atmosphere of the planet (top two panels) and in the disc (bottom two panels) in the inheritance scenario with high ionisation. The top two panels show results for the six migration scenarios, corresponding to initial semimajor axes of 5, 12, 19, 50, 100, and 130 au. The increasing size of the markers maps the increasing distance travelled by the planet in the six migration scenarios. Linking lines are shown only for visual ease. The plot on the left (with empty markers) describes a gas-dominated giant planet, while the plot on the right (filled markers) is for solid-enriched ones. The bottom four panels show the trends of the elemental ratios on the disc for the gas (dashed lines) and the solid (solid lines) phases. The black, dashed, horizontal line at 1 indicates the reference stellar value in all plots.}
	\label{fig:ternary2}
\end{figure*}

\begin{figure*}
	\includegraphics[width = \textwidth]{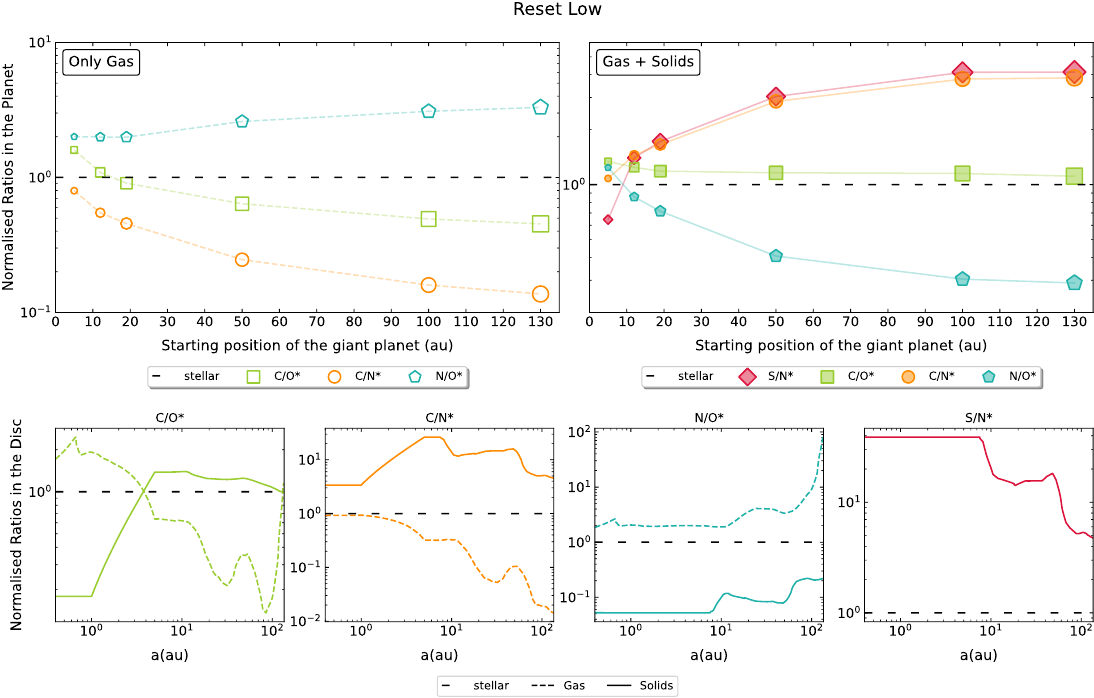}
	\caption{Normalised elemental ratios in the final atmosphere of the planet (top two panels) and in the disc (bottom four panels) in the reset scenario with low ionisation. The top two panels show results for the six migration scenarios, corresponding to initial semimajor axes of 5, 12, 19, 50, 100, and 130 au. The increasing size of the markers maps the increasing distance travelled by the planet in the six migration scenarios. Linking lines are shown only for visual ease. The plot on the left (with empty markers) describes a gas-dominated giant planet, while the plot on the right (filled markers) is for solid-enriched ones. The bottom four panels show the trends of the elemental ratios on the disc for the gas (dashed lines) and the solid (solid lines) phases. The black, dashed, horizontal line at 1 indicates the reference stellar value in all plots.}
	\label{fig:ternary3}
\end{figure*}

\begin{figure*}
	\includegraphics[width = \textwidth]{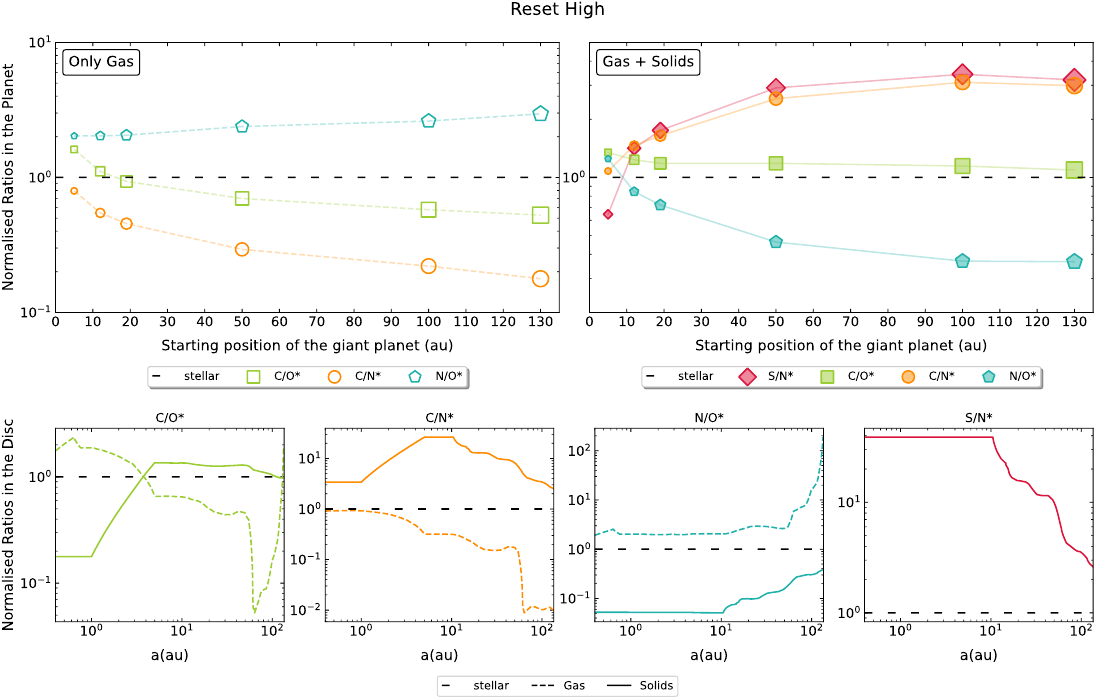}
	\caption{Normalised elemental ratios in the final atmosphere of the planet (top two panels) and in the disc (bottom four panels) in the reset scenario with high ionisation. The top two panels show results for the six migration scenarios, corresponding to initial semimajor axes of 5, 12, 19, 50, 100, and 130 au. The increasing size of the markers maps the increasing distance travelled by the planet in the six migration scenarios. Linking lines are shown only for visual ease. The plot on the left (with empty markers) describes a gas-dominated giant planet, while the plot on the right (filled markers) is for solid-enriched ones. The bottom four panels show the trends of the elemental ratios on the disc for the gas (dashed lines) and the solid (solid lines) phases. The black, dashed, horizontal line at 1 indicates the reference stellar value in all plots.}
	\label{fig:ternary4}
\end{figure*}

\begin{figure*}
	\includegraphics[width = \textwidth]{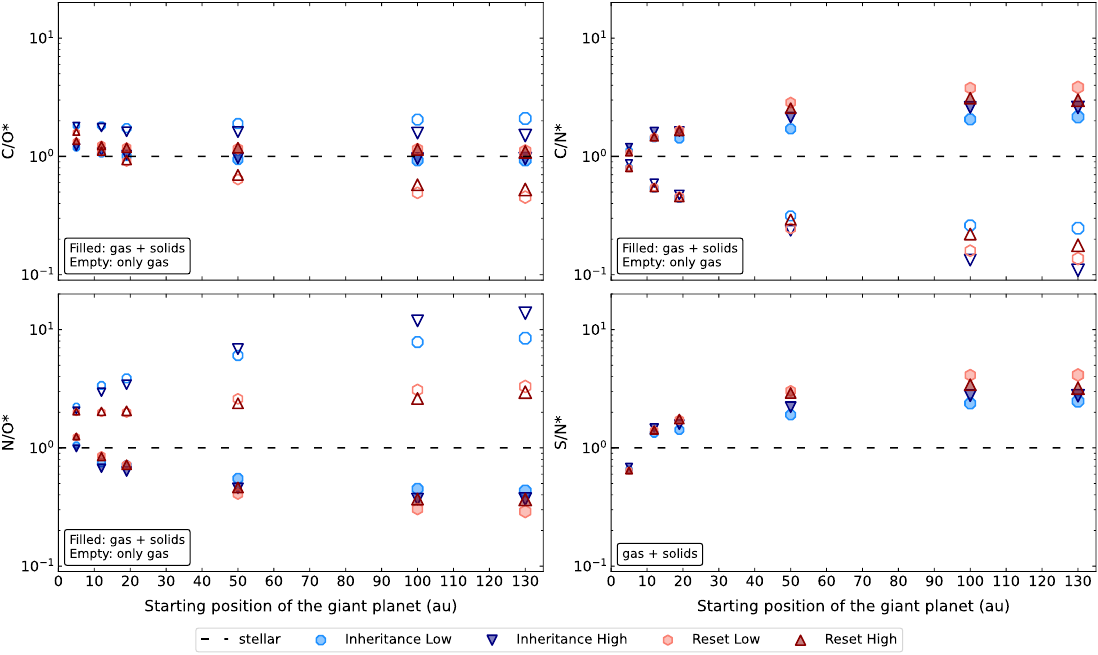}
	\caption{Elemental ratios C/O*, C/N*, N/O*, and S/N* normalised to their respective stellar value in the four chemical scenarios. The data are the same as in the top two panels of Figures~\ref{fig:ternary1}, \ref{fig:ternary2}, \ref{fig:ternary3}, and \ref{fig:ternary4}, here being plotted together to be more easily compared with each other. The blue points refer to the inheritance scenarios, while the red ones refer to the reset scenarios. Specifically, the light-blue octagons and the light-red hexagons indicate low ionisation level, while the dark-blue, downward triangles and the dark-red, upward triangles indicate high ionisation level. The results are plotted for the six migration scenarios, corresponding to initial semimajor axes of 5, 12, 19, 50, 100, and 130 au. The increasing size of the markers maps the increasing distance travelled by the planet in the six migration scenarios. Trends for both the accretion scenarios are shown: filled markers for solid-enriched and empty markers for gas-dominated giant planets. The black, dashed, horizontal line at 1 indicates the reference value for giant planets whose chemical composition matches that of their host star.}
\label{fig:comparison}
\end{figure*}

The results of our analysis are plotted for each of the four scenarios in Figures~\ref{fig:ternary1}, \ref{fig:ternary2}, \ref{fig:ternary3}, and \ref{fig:ternary4}, while their trends for each elemental ratio are compared in Figure~\ref{fig:comparison}. The results reveal that the joint use of different elemental ratios proves to be the key to accurately characterise the formation of giant planets in terms of:
	\begin{itemize}
		\item the accretion history, i.e., whether it was dominated by the gas or the planetesimals,
		\item the extent of the migration, 
		\item the chemical initial conditions of the protoplanetary disc (only for gas-dominated giant planets).
	\end{itemize}
We will delve into each of these aspects in Sections~\ref{sec:res_accretion}, \ref{sec:res_migration}, and \ref{sec:res_chemistry}. Before doing that, however, it is worth discussing some important concepts to have in mind when interpreting the information in Figures~\ref{fig:ternary1}, \ref{fig:ternary2}, \ref{fig:ternary3}, and \ref{fig:ternary4}.  

%%% ######################################################################## %%%
\subsection{Linking planetary atmospheres to disc structures}\label{sec:ternary_guide} 

The final composition of giant planets is generally expected to reflect the chemical structure of the protoplanetary disc, especially that of the region of the disc where the planet was born \citep[e.g.,][]{oberg2011,lothringer2021}. We can assess the degree to which this assumption is reasonable by comparing the elemental ratios in the two environments. The results of such analysis are illustrated in Figures~\ref{fig:ternary1}, \ref{fig:ternary2}, \ref{fig:ternary3}, and \ref{fig:ternary4}. Each figure examines one of the four chemical scenarios. The top two panels show the normalised elemental ratios in the atmosphere of gas-dominated (on the left) and solid-enriched (on the right) giant planets at the end of the six simulations. The lines connecting the different scenarios are provided as a visual aid only. The bottom four panels show the radial profiles of the normalised elemental ratios in the disc, as computed from Figure~\ref{fig:disc_comp_elem}. Such profiles can be seen as snapshots of the midplane chemical structure, where solid and dashed lines characterise the solid and gas phase, respectively. 

Overall, the four figures show that the atmosphere of the giant planet inherits only part of the chemical properties of the disc. Specifically, elemental ratios that are sub- or superstellar in the disc will generally be so in the planet as well. However, the results also show that this disc-planet link cannot be used to make quantitative predictions. Specifically, the plots reveal that for some migration scenarios the normalised elemental ratios in the planet and in the disc do not immediately relate. For instance, although the S/N* is markedly superstellar throughout the disc and increases towards the star, its value in the solid-enriched giant planet that starts forming at 5 au drops to the substellar level. The final composition of the planet envelope depends crucially on how efficiently the gas and the solids are accreted from each region of the disc along the migration pathway. For gas-dominated giant planets, the accretion of gas reaches its maximum during the runaway gas accretion phase (see Section~\ref{sec:formmodel}). The final atmosphere of the planet will therefore be strongly influenced by the chemical properties of the region of the disc traversed by the planet during this phase. In solid-enriched giant planets the accretion of gas and solids are coupled, but their relative contribution to the planetary metallicity depends on the extent of the migration. In particular, the larger the migration the larger the number of planetesimals encountered by the planet and the larger the contribution of planetesimal accretion to the final metallicity. As the planet migrates for shorter distances, it encounters and accretes fewer planetesimals and the contributions to its metallicity by gas and solids become comparable. As a consequence, the elemental ratios of solid-enriched giant planets follow the global trends of the disc solid component only in large-scale migration scenarios, while diverging from them for limited migrations.

The decreasing trend of the planetary S/N* ratio from large- to short-scale migrations can be easily understood considering how S and N are distributed in the disc and how they are accreted on the planet. In our disc, S is entirely locked in the solid phase, while N is almost entirely in the gas phase as N$_2$, slightly deviating from the stellar value beyond the NH$_3$ snowline (see Figure~\ref{fig:disc_comp_elem}). As a direct consequence of this partitioning, in the planet envelope the total S scales linearly with the accreted planetesimals, while the total N is accreted in stellar proportion almost independently on the radial migration. Consequently, even though the S/N* ratio of solids in the inner disc is about 40 times higher than the stellar value, for short-scale migrations the contribution of the accreted N becomes large enough to bring the S/N* ratio in the planet to substellar value. Similar arguments can be used to explain why the N/O*, C/O*, and C/N* ratios of solid-enriched giant planets approaches stellar values for short-scale migrations. 

Our results emphasise how the composition of giant planets and their atmospheres can be connected to that of their native discs only by coupling the disc chemical structure with the growth and migration tracks that characterise planet formation. With this in mind, we can proceed with the discussion of our results and their implications for planet formation.    
 
%%% ######################################################################## %%%
\subsection{The diagnostic power of the C/O ratio}\label{sec:coratio} 

Figure~\ref{fig:comparison} compares the normalised elemental ratios of solid- and gas-dominated giant planets in the inheritance (blue points) and reset (red points) scenarios. The results are shown for each of the six simulations, identified by the location of the planet at the onset of its formation. The C/O* ratio shows almost flat trends, revealing a rather limited diagnostic power.

For short-scale migrations, the C/O* ratios in all scenarios vary between 1 and 2 times the stellar value. The limited variations between the different disc chemistry and planet migration scenarios are too small to be unequivocally resolved with the current accuracy of retrieval methods, which is of the order of $20\%$ (see \citealt{barstow2020} and Paper I for further discussion). Therefore, giant planets that formed close to the star in a disc that inherited its composition from the pre-stellar core cannot be observationally distinguished from those that formed in a disc that experienced a complete reset of the chemistry, nor can we resolve whether they are gas dominated or solid enriched.

The degeneracy is partially broken for large-scale migrations, where the C/O* ratio of solid-enriched and gas-dominated giant planets follows distinct trends. Specifically, solid-enriched giant planets are characterised by stellar values of the C/O* ratio, independent of how far the planet started its migration and the chemical structure of the birth environment. On the contrary, gas-dominated giant planets are characterised by a C/O* ratio significantly deviating from the stellar value. In this case, the C/O* ratio provides constraints on both the migration and the disc chemical scenario. Detailed discussion on each of these individual aspects is provided in Sections~\ref{sec:res_accretion}, \ref{sec:res_migration}, and \ref{sec:res_chemistry}.

%%% ######################################################################## %%%
\subsection{Constraints on the accretion history}\label{sec:res_accretion} 

As anticipated in Section \ref{sec:coratio}, the C/O* ratio provides information on the accretion history of giant planets only for large-scale migration scenarios. In this case, stellar values are associated with solid-enriched giant planets, whereas marked sub- and superstellar values provide an indication of accretion dominated by the gas. Specifically, substellar C/O* ratios are associated with gas-dominated giant planets in the reset scenarios, whereas superstellar values characterise gas-dominated giant planets in the inheritance scenarios. 

To understand the origin of such trends we refer to the disc composition, as described by the molecular abundance profiles in Figure~\ref{fig:profiles}. Beyond 10 au, the gas phase in the reset scenarios is characterised by a lower abundance of CH$_4$ and higher abundances of CO and O$_2$ with respect to the inheritance scenarios. Because both CO and O$_2$ typically condense at very low temperatures, they act as reservoirs of gaseous O in most of the disc extension. Consequently, the abundance of total O* in the gas phase is higher in the reset scenarios than in the inheritance ones. In particular, beyond 10 au, the abundance of O* exceeds that of C*. Such behaviour is shown in Figure~\ref{fig:disc_comp_elem} and is the reason why the C/O* ratio of the disc gas phase drops below the stellar value in the reset scenarios (see Figures~\ref{fig:ternary3} and \ref{fig:ternary4}). As higher abundances in the gas correspond to lower abundances in the solids, the opposite trend is observed for the C/O* ratios of the disc solid phase. Moving from the disc to the planet, we recall that the final elemental ratios in the planet envelope depend on the mass accretion rate (see the discussion in Section~\ref{sec:ternary_guide}). Such rate is a non-linear function of the orbital distance, and it is highly influenced by the position on the disc where the runaway gas accretion occurs. As discussed in Section~\ref{sec:formmodel}, in our model the runaway gas accretion phase begins once the first 40$\%$ of the total radial displacement of the planet has been covered. For migrations starting beyond 20 au, in our simulations, this means that the region traversed by the planet during the runaway gas accretion phase falls between 10 au and 80 au. Therefore, gas-dominated giant planets in the reset scenarios accrete most of their gas from the disc region in which the C/O* ratio is substellar, which is why their final C/O* ratio is also substellar. When it comes to solid-enriched giant planets, the contribution of planetesimal accretion to the envelope metallicity increases with the extent of migration. For large-scale migrations, the metallicity is dominated by the planetesimals accreted from the disc region in which the C/O* ratio of the solids is superstellar, which is why the final C/O* ratio in the planet envelope is also superstellar.

We highlight that in the inheritance scenario with a high ionisation level, the C/O* ratio of gas-dominated giant planets slightly decreases with increasing length of migration. For the accuracy of the current retrieval methods (see Section \ref{sec:coratio}), this introduces a degeneracy of the planet's accretion history. Specifically, for very large-scale migration scenarios, gas-dominated giant planets in the inheritance high scenario would be indistinguishable from solid-enriched ones. Such a decreasing trend in the inheritance high scenario originates from the enhanced abundance of O* between 50 au and 100 au in our disc model (see Figure~\ref{fig:disc_comp_elem}), which in turn results from the peak abundance of O$_2$ in the outer disc (see Figure~\ref{fig:profiles}). In this region, the abundance of O* exceeds that of C*, hence bringing the disc C/O* ratio to substellar values (see Figure~\ref{fig:ternary2}). Gas-dominated giant planets that accrete most of their gas from this region will therefore be characterised by lower C/O* ratios than in the inheritance low scenario. 

The information provided by the C/O* ratio on the planet accretion history can find independent confirmation in the total metallicity of the planet envelope. Gas-dominated giant planets are characterised by substellar envelope metallicities that decrease with migration, as highlighted also in Paper I. Solid-enriched giant planets are instead characterised by the opposite trend, i.e., superstellar envelope metallicities that increase with migration. Note that here the relevant quantity is the planetary metallicity normalised to that of the host star. The choice of the right stellar reference value is therefore key to the correct interpretation of observational data. The C/O ratio and metallicity of the Sun are widely used, either explicitly or implicitly, as referenced in interpreting the composition of exoplanetary atmospheres (see, e.g., \citet{madhusudhan2019} for a recent review). However, both the C/O ratio and the metallicity of stars in planetary systems can significantly deviate from their respective values in the Sun (e.g., \citealp{delgado2010, mulders2018} and references therein; \citealp{magrini2022}). Therefore, to avoid introducing biases, one has to look at the planetary elemental ratios and metallicity in the reference frame of the host star. For instance, a planetary C/O ratio estimated to be 0.55 (equal to the solar C/O ratio) would be stellar only if the planet formed around a solar-type star. If, instead, the planet formed around a star characterised by a subsolar C/O ratio (e.g., 0.45), the planetary C/O ratio should be correctly interpreted as superstellar. Our methodology has been recently successfully applied by \citet{carleo2022} and \citet{guilluy2022} to the interpretation of observational data from the GAPS 2 program.

It is important to notice that when the C/O* ratio can only be estimated as being either sub- or superstellar, it is not possible to constrain the planet's accretion history, unless independent measurements of the envelope metallicity are available. 

Our findings reveal that using the C/O* ratio alone to probe the formation pathways of giant planets could lead to wrong conclusions. The limited diagnostic power of the C/O* ratio is essentially due to the high volatility of both C and O, which makes the ratio unable to unequivocally trace the accretion of gas and solids. The inclusion of N, one of the most volatile elements, in the set of elemental tracers allows the computing of the C/N* and N/O* ratios and the breaking of the degeneracy on the accretion history. Specifically, Figure~\ref{fig:comparison} shows that for both the C/N* and N/O* ratios the trends for gas-dominated and solid-enriched giant planets are well separated and always characterised by opposite behaviours. Such behaviours result from the lower volatility of C and O with respect to N. Specifically, while the bulk of N in discs remains in the gas phase as N$_2$, the abundances of gaseous C and O decrease towards the outer disc (see Figure~\ref{fig:disc_comp_elem}). Variations of the C/N* and N/O* ratios in discs are therefore driven by variations of C* and O*. As a consequence, the C/N* (N/O*) ratio of gas-dominated giant planets is globally substellar (superstellar) and decreases (increases) with the radial migration. The behaviour of solid-enriched giant planets is the opposite, with C/N* (N/O*) globally superstellar (substellar) and increasing (decreasing) with migration.  

It is worth noticing that the N/O* ratio of gas-dominated giant planets is always markedly superstellar and never drops below 2. Such a trend is due to the fact that in our model about 50$\%$ of O  is early sequestered into rocks, hence subtracted from the gas phase. As a consequence, in short-scale migration scenarios, even if the other elemental ratios approach the stellar value, the N/O* ratio does not. 

What emerges from our study is that in order to get a complete picture of planet formation, one has to look at multiple elemental ratios. One of the advantages of using normalised ratios is that their trends can be more readily and intuitively compared. The comparison of the top two panels in Figures~\ref{fig:ternary1}, \ref{fig:ternary2}, \ref{fig:ternary3}, and \ref{fig:ternary4} reveals the existence of fixed relations between the ratios of gas-dominated and solid-enriched giant planets. In particular, regardless of the chemical scenario, the N/O* ratio of gas-dominated giant planets is always larger than the C/O* ratio, which is in turn larger than the C/N* ratio. The opposite relation holds for solid-enriched giant planets for moderate and large-scale migrations. Such relations result from the differences in relative volatility between C, O, and N. More volatile elements are characterised by lower condensation temperatures. As such, they condense in solid form farther away from the star and remain in gas form over wider disc regions. As discussed in Section~\ref{sec:compmodel}, in our compositional model, N is almost entirely present in the gas phase as N$_2$ in all four scenarios. Therefore, as a result of its high volatility, N will be proportionally more abundant than C and O in giant planets that accreted only gas, hence leading to N/O*$>$C/O*$>$C/N*. For solid-enriched planets, which accreted from the solid phase that is progressively depleted in highly volatile elements, the opposite is true and C/N*$>$C/O*$>$N/O*. The emerging picture is that a joint evaluation of normalised elemental ratios allows us to immediately and unequivocally discriminate between planets that derive their metallicity from the accretion of gas and planets that experienced planetesimal enrichment. These results expand those of Paper I, showing how the relative values of the normalised ratios follow the same patterns independently on the underlying chemical scenario. These behaviours have been recently confirmed in the studies of \citet{biazzo2022}, \citet{carleo2022}, and \citet{guilluy2022}, based on Paper~I. 

%%% ######################################################################## %%%
\subsection{How far did the planet start its migration?}\label{sec:res_migration} 

Among all the ratios, C/O* appears to be the least robust as a diagnostic tool for the migration history of the planet. In particular, Figure~\ref{fig:comparison} shows that the C/O* ratio exhibits limited changes among the migration pathways in almost all chemical scenarios. Its overall flatness is essentially due to the limited range of variation of the C/O* ratio in the disc with respect to the other ratios, as it is shown in the bottom four panels of Figures~\ref{fig:ternary1}, \ref{fig:ternary2}, \ref{fig:ternary3}, and \ref{fig:ternary4} (note the different scales on the $y$-axes). The strongest effects on the planetary C/O* ratio are observed in the two reset scenarios for gas-dominated giant planets, where the values decrease by a factor of 3 (i.e., from 1.6 to 0.5), for formation regions increasingly farther away from the star. In particular, the trend decreases steadily up to 50 au and progressively flattens further out. We discussed the origin of such a trend in Section \ref{sec:res_accretion}. On the contrary, for both gas-dominated and solid-enriched giant planets, the C/N* and N/O* ratios are significantly more sensitive to the radial migration than the C/O* ratio. Specifically, they show monotonic trends with deviations from the stellar value that increase with direct dependency on the migration, leading to separations up to one order of magnitude between short- and large- scale migrations. 

\begin{figure}
	\centering
    \includegraphics[width=\hsize]{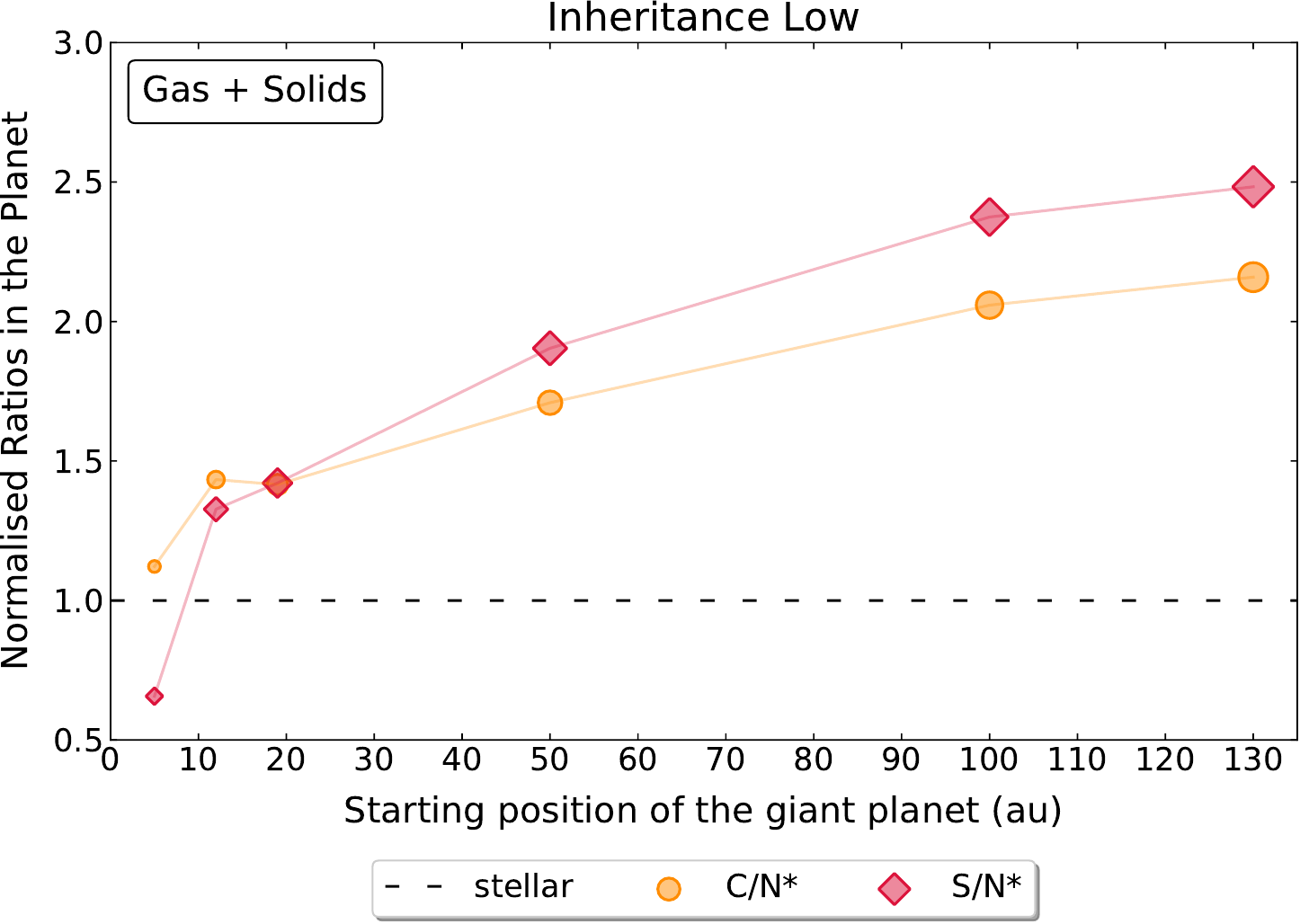}
	\caption{Comparison between the C/N* and the S/N* ratios of solid-enriched giant planets in the inheritance low scenario. The trends are taken from Figure~\ref{fig:ternary1} and shown in linear scale to highlight the differences. The black, dashed, horizontal line at 1 indicates the reference stellar value.}
	\label{fig:cnsnlin}
\end{figure}

The S/N* ratio provides additional information for solid-enriched giant planets. As it is shown in the top-right panel of Figures~\ref{fig:ternary1}, \ref{fig:ternary2}, \ref{fig:ternary3}, and \ref{fig:ternary4}, and highlighted in Figure~\ref{fig:cnsnlin}, large-scale migrations of solid-enriched giant planets result in S/N* ratios systematically above C/N* ratios. Although the difference between the two may appear marginal, the trend results from the lower volatility of S with respect to C, which causes a larger fraction of S to be locked in the solid phase than C. Given that the total budget of S and C is dominated by the accretion of solids for large-scale migrations, the abundance of S grows faster than that of C, resulting in S/N*$>$C/N*. On the contrary, short-scale migrations of solid-enriched giant planets result in S/N* ratios markedly lower than C/N* ratios. Such a trend is a consequence of the increasingly dominant contribution of the gas accretion (see the discussion in Section \ref{sec:ternary_guide}) combined with the fact that the gas is the major carrier of C within the snowline of refractory organic carbon. Our results confirm that the relation between the S/N* and C/N* ratios holds for all the disc chemical scenarios in which the bulk of N remains in gaseous form while the bulk of S is early trapped in refractory material, so that the planetary S/N* ratio becomes a direct tracer of the planetesimal accretion. 

Given its intrinsic properties, the S/N* ratio is an extremely versatile diagnostic tool. In particular, the joint evaluation of the S/N* ratio and the envelope metallicity of the planet allows discrimination between two scenarios that are not directly modelled by our simulations. The first one describes a giant planet that formed in situ and close to the star, in a hot environment with $T>700$~K. At such a high temperature, all the volatile elements, including S, are in the gas phase. In cosmochemistry, S is indeed the most abundant among the moderately volatile elements \citep{palme2014}, with condensation temperature $T_{\text{cond}}= 672$~K \citep{wood2019}. Consequently, the planet would be characterised by stellar values of both the envelope metallicity and the S/N* ratio. The second scenario is that of a solid-enriched hot Jupiter that accreted most of its heavy-element budget beyond the N$_2$ snowline, where even the highly volatile molecules (CO and N$_2$) are condensed in the solid phase \citep{oberg2019}. At the end of its migration, when the planet reaches its close-in orbit, its envelope would be characterised by a markedly superstellar metallicity and a stellar S/N* ratio. Therefore, the joint evaluation of the metallicity and the S/N* ratio allows distinguishing hot Jupiters that formed in situ and close to the star from those that underwent extreme migrations. 

The combined use of metallicity, C/N*, and S/N* ratios allows the investigation of another alternative scenario that was not directly modelled by our simulations. This is the case of a giant planet that formed in a disc in which the inward drift of dust and pebbles causes the ices of C, O, and N to sublimate as they cross their respective snowlines \citep{booth2019, schneider2021}, thus directly enriching the disc gas phase in heavy elements. In such scenario, gas accretion may then contribute more than solid accretion to the total metallicity of the planet, hence mimicking the effect of short-scale migration in our simulations. Due to the higher volatility of C with respect to S, sublimation would be more efficient in enriching the gas in C rather than in S. Therefore, hot Jupiters that form in such scenario, will be characterised by superstellar metallicity and C/N*$>$S/N*.

%%% ######################################################################## %%%
\subsection{The chemical structure of the birth environment}\label{sec:res_chemistry} 

As discussed in Sections~\ref{sec:res_accretion} and \ref{sec:res_migration}, our new analysis confirms and expands the general conclusion of Paper~I about the limited predictive power of the C/O* ratio. Nevertheless, the C/O* ratio does retain some diagnostic power in our model. Specifically, the C/O* ratio of gas-dominated giant planets that underwent moderate and large-scale migrations (beyond 10~au in our model) is always superstellar in the inheritance scenarios and substellar in the reset ones (see the top-left panel in Figure~\ref{fig:comparison}). Therefore, provided that it is possible to independently unveil the origin of the accreted material, e.g., by looking at the other elemental ratios or through density and metallicity estimates, the C/O* ratio of gas-dominated giant planets can be used to constrain the chemical structure of the birth environment. On the other hand, little can be said about solid-enriched giant planets, whose C/O* ratio does not deviate appreciably from the stellar value in both scenarios.

For gas-dominated giant planets, the N/O* ratio in the atmosphere provides additional insights into the chemical characterisation of the disc. As it is shown in the bottom-left panel of Figure~\ref{fig:comparison}, the results for the inheritance and the reset scenarios follow two distinct trends that diverge with respect to each other. The separation between the two increases with the radial migration, resulting in a N/O* ratio systematically higher in the inheritance scenarios than in the reset ones. Specifically, for large-scale migrations, the N/O* ratio in the inheritance scenarios can be more than a factor of three larger than in the reset ones. Such a trend is a direct consequence of the higher abundance of gaseous O in the outer disc in the reset scenarios with respect to the inheritance ones (see Figure~\ref{fig:disc_comp_elem}). We discussed this feature in Section~\ref{sec:res_accretion} to explain why the C/O* ratio of gas-dominated giant planets is substellar in the reset scenarios. 

%%% ######################################################################## %%%
\subsection{Pairwise comparison of elemental ratios}	

\begin{figure*}
	\includegraphics[width = \textwidth]{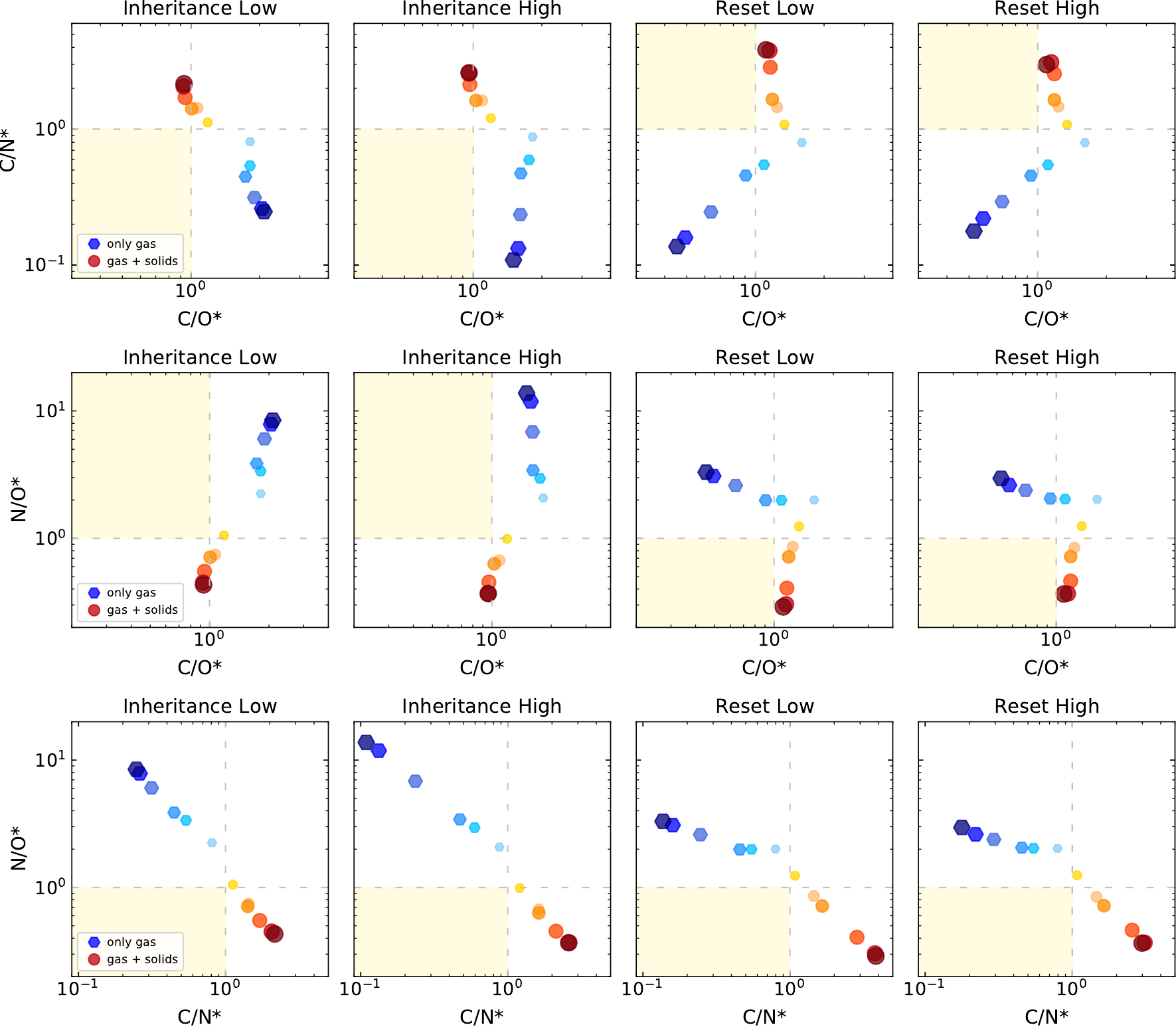}
	\caption{Comparison between pairs of normalised elemental ratios in the planet's atmosphere. Results are shown for the four analysed chemical scenarios. Each point on the plots represents one of the six analysed simulations, describing a planet that starts its migration at 5, 12, 19, 50, 100, and 130 au from the star. Colour maps and increasing size of markers are used to distinguish between short-scale (light colours, small markers) and large-scale (dark colours, big markers) migrations. The blue markers (hexagons) indicate gas-dominated giant planets, while the red ones (circles) are for solid-enriched ones. The grey, dashed lines indicate stellar values. Their intersection identifies four quadrants in which the elemental ratios can assume sub- or superstellar values. The yellow shaded areas indicate regions of the parameter space that are forbidden in our simulations.   
	}
	\label{fig:binary}
\end{figure*}

To provide a more immediate visualisation of the characteristics of planetary atmospheres and to facilitate the interpretation of observational data, we compared pairs of normalised elemental ratios through binary plots. Figure~\ref{fig:binary} illustrates this comparison for three pairs of ratios, C/N* versus C/O*, N/O* versus C/O*, and N/O* versus C/N* for both gas-dominated and solid-enriched giant planets in the four chemical scenarios. The grey dashed lines, indicating stellar values, delimit four quadrants in which the ratios can assume sub- or superstellar values. One of the advantages of this representation is that the results for different combinations of the parameters of the system (i.e., the extent of migration, the accretion, and chemical scenarios) occupy distinct quadrants. One can then immediately identify the conditions that favour one scenario over the others. 

For instance, a point in the third quadrant (bottom left) of the C/N* versus C/O* plot, i.e., substellar values for both the C/N* and C/O* ratios, identifies a planet that started its migration far from the star (beyond the CO$_2$ snowline) and accreted only gas in a reset scenario. Conversely, a superstellar C/O* ratio coupled with a C/N* ratio an order of magnitude lower than the stellar value indicates that the same formation process occurred in an inheritance scenario.

Moreover, the comparison between the N/O* and C/N* ratios allows for constraining the accretion history and the migration of giant planets. Specifically, the deviations of the ratios from their respective stellar value are directly proportional to the radial migration and grow in opposite directions for gas-dominated and solid-enriched giant planets. In the inheritance scenarios (see the two bottom-left panels), the N/O* and C/N* ratios of gas-dominated giant planets can deviate up to one order of magnitude above and below the stellar value, respectively. For the same planets, moderate deviations of the N/O* ratio with respect to the C/N* ratio provide a strong indication that the evolution occurred in a reset scenario (see the two bottom-right panels). When considering solid-enriched giant planets, both the N/O* and C/N* ratios are less sensitive to migration than in gas-dominated planets and follow the same trend independently on the chemical scenario. 

Alongside favoured regions of the parameter space for the different planet formation histories, our results suggest forbidden (as indicated by the yellow shaded areas in Figure~\ref{fig:binary}) or marginally permitted quadrants in pairwise elemental ratios analysis. For instance, in no case among those we investigated in our modelling, can planets populate the quadrant where both the N/O* and the C/N* ratios are substellar (see the bottom four panels of Figure~\ref{fig:binary}). The opposite case, i.e., N/O* and C/N* ratios simultaneously superstellar, is  very marginally compatible only with solid-enriched giant planets that experienced very short-scale migrations, starting within the H$_2$O snowline.

%%% ######################################################################## %%%
\subsection{Refractory elements beyond S}

One of the main findings of Paper I, confirmed by this study, is the untapped potential of the S/N* ratio as a diagnostic tool for planet formation. As discussed in Section \ref{sec:res_migration}, this is essentially due to the marked volatility contrast between S and N, which makes their ratio an effective tracer of the accretion of solids. It is important to point out, however, that any element X more refractory than O can be used in place of S to perform the same analysis. From an observational perspective, this means that our set of diagnostic tracers can be extended to elements that may be more easily detectable than others in giant planet atmospheres. For instance, for increasing equilibrium temperatures of the planets, the atmospheres get enriched in elements with increasingly lower volatility. For warm and hot Jupiters, moderately volatile elements are all good alternatives to S. Such category includes elements that condense at temperatures between 1250~K and 250~K, such as Na, K, Cl, and F \citep{palme2014, wood2019}. For ultrahot Jupiters the list extends to refractory (e.g., Al, Ca, and Ti) and rock-forming (e.g., Mg, Si, Ni, and Fe) elements, which have condensation temperatures in the range 1850-1250~K \citep{palme2014, wood2019}. 

In Section~\ref{sec:res_migration} we stressed that the planetary elemental ratios strongly depend on the phase of the main carrier of the two involved elements (e.g., the gas phase for N$_2$, which is the main carrier of N and the solid phase for the rocks that are the main reservoir of S) and on how efficiently the gas and the solids contribute to the final envelope metallicity. In particular, we examined the case of the C/N* and S/N* ratios in solid-enriched giant planets. Which of the two ratios is larger than the other depends on whether the planet's formation is dominated by the accretion of gas or solids. More generally, the comparison between two elemental ratios X/N* and Y/N*, where the elements X and Y have a marked volatility contrast, allows for discrimination between scenarios in which one of the two elements is predominantly accreted by different phases.

In Section \ref{sec:res_migration} we also highlighted that the S/N* ratio allows for resolving scenarios of in situ formation very close to the star, in a hotter environment than the condensation temperature of S. Specifically, giant planets that formed by accreting gas from such a region will be characterised by stellar values of both metallicity and S/N* ratio. It is important to notice that the replacement of S with a more refractory element would push this limit towards higher temperatures, allowing us to resolve formation scenarios even closer to the star.

In Section~\ref{sec:res_migration} we also discussed how the comparison between the metallicity and the C/N* and S/N* ratios allows the identification of giant planets that formed by accreting gas enriched in heavy elements due to the sublimation of inwardly drifting dust and pebbles, contrary to planetesimal accretion. Sublimation is more efficient in enriching giant planets in highly volatile elements rather than in refractory ones. In particular, any refractory element X would produce less enrichment in the gas than in C. Therefore, planets that formed by accreting high-metallicity gas would be characterised by superstellar metallicity and C/N*$>$X/N*.

%%% ######################################################################## %%%
\subsection{Linking our predictions to observable planetary atmospheres}

In the previous sections we discussed how the atmospheric elemental ratios can be used to trace the formation history of giant planets. From an observational perspective, such elemental ratios need to be retrieved from the abundances of the molecular carriers of C, O, N, and S in exoplanetary atmospheres.  

The observable chemistry in giant planet atmospheres is shaped by multiple and diverse processes, such as thermochemical reactions at equilibrium, mixing processes, photochemistry, and chemical diffusion. Each of them dominates in different atmospheric layers (e.g., \citealp{madhusudhan2016} and references therein). 

Deep in the atmosphere, the chemistry is at equilibrium and the chemical composition is a function only of the elemental abundances, the temperature, and the pressure. For a solar composition, the main molecular carriers of C and N are CO, CH$_4$, N$_2$, and NH$_3$. Two regimes in the pressure-temperature space can be identified. At high temperatures and low pressures, C and N are essentially locked in CO and N$_2$, while at low temperatures and high pressures, the main reservoirs of C and N are CH$_4$, and NH$_3$. At the pressure of 1 bar, the transition between the two regimes occurs at the temperatures of 1200 K  and 700 K for the C- and N- bearing molecules, respectively. O is largely present as water vapour, while S is mainly locked into H$_2$S (e.g., \citealp{madhusudhan2016}, \citealp{fortney2021} and references therein). The chemical equilibrium in atmospheres with supersolar metallicity generally favours the production of molecules that contain two or more heavy elements. In such atmospheres, CO$_2$ is therefore favoured over CO and N$_2$, which in turn are favoured over CH$_4$ and NH$_3$ (e.g., \citealp{madhusudhan2016} and references therein). 
 
In the upper layers of the atmosphere, photochemical reactions induced by stellar UV radiation are proved to alter the chemical composition. Among the direct and indirect products of photochemistry there are H, HCN, C$_2$H$_2$, OH, O$_2$, and NO. Regarding S, photochemistry converts H$_2$S in S, HS, S$_2$, SO, and SO$_2$ (e.g., \citealp{madhusudhan2016, fortney2021}). 

Chemistry in observed exoplanetary atmospheres may be more complex than discussed above, being also influenced by the cooling history of the planet \citep{fortney2020}. At intermediate layers, volatile species may be sequestered into clouds, although these are not expected to form at the high atmospheric temperatures of hot Jupiters (e.g., \citealp{madhusudhan2016}, \citealp{madhusudhan2019} and references therein). 

Ground-based and high-resolution spectral observations have recently succeeded in providing the first characterisations of exoplanetary atmospheres. For instance, observations by \citet{giacobbe2021} of the hot Jupiter HD209458b, at the equilibrium temperature of 1500 K, revealed the presence of H$_2$O, CO, HCN, CH$_4$, NH$_3$, and C$_2$H$_2$ in the atmosphere. \citet{carleo2022} and \citet{guilluy2022} have recently extended this result to giant planets with equilibrium temperature $T\leq 1000$K. Future observations with JWST \citep{greene2016}, Twinkle \citep{edwards2019}, and Ariel \citep{tinetti2018}, as well as advances in atmospheric retrieval techniques, will soon revolutionise the field. The broad spectral range and the high spectral resolution of JWST will allow for precise determination of the abundances of a number of species of interest besides H$_2$O, like CO, CO$_2$, CH$_4$, and NH$_3$ \citep{greene2016}. An even greater number of molecules will be accessible with Ariel, by the time the telescope starts its operations in 2029 \citep{tinetti2018}. 

Recent studies have validated our methodology on real data in preparation of its use for interpreting the observations of future ground-based and space-based facilities. Alongside \citet{carleo2022} and \citet{guilluy2022}, \citet{biazzo2022} put constraints on possible formation scenarios for the giant planets in four planetary systems, around the stars HAT-P-26, WASP-39, HAT-P-12, and WASP-10. Specifically, they analysed the trends of the normalised planetary metallicity and elemental ratios based on the atmospheric composition obtained by \citet{kawashima2021} through spectral disequilibrium retrieval models. A similar study was performed independently by \citet{kolecki2022} on an additional set of four giant planets. As discussed before, \citet{carleo2022} and \citet{guilluy2022} jointly used the planetary metallicity and the C/O* ratio to constrain the formation histories of WASP-69b and WASP-80b, as well as the characteristics of their native circumstellar discs (see also Section~\ref{sec:res_accretion}). More systematic studies of the observational implications of our results will be the focus of dedicated future works. 

%%% ######################################################################## %%%

\section{Conclusions}\label{sec:concl}

In this work, we investigated the link between the final composition of the atmospheres of giant planets and their formation process, focusing on the effects of different chemical initial conditions for the host protoplanetary disc. The study in Paper I was limited to a disc that inherited its composition from the pre-stellar core and whose chemical evolution was limitedly affected by cosmic rays and similar energy sources. Here we also explored the case of a complete reset of the chemistry, as well as the impact of different levels of the ionisation rate throughout the disc. Specifically, we considered a first case in which the only source of ionisation is the decay of short-lived radionuclides (i.e., the same considered in Paper I) and a second case in which an additional contribution is provided by cosmic rays. In addition, we introduced a more realistic model of the condensation of refractory organic carbon. 

We analysed the six $N$-body simulations from Paper~I of the growth and migration of a giant planet in a disc of gas and planetesimals. During the migration, the planet grows in radius and mass by accreting gas and solids characterised by different compositions and relative abundances of refractory and volatile elements. By coupling the outcome of the simulations with the different compositional models of the disc we derived the composition of the accreted material, from which we computed the C/O, N/O, C/N, and S/N ratios in the final atmosphere of the planet. For a better interpretation of the results, we normalised the elemental ratios to their corresponding stellar values and we indicated the new values with the superscript *. Our findings can be summarised as follows: 
 \begin{itemize}
 
 	\item The final composition of giant planets is markedly non-stellar due to the interplay between planetary migration and concurrent accretion of gas and solids. Therefore, despite being intimately connected, the planetary elemental ratios are not a direct reflection of those of the disc.
 	
 	\item The diagnostic power of the elemental ratios is maximised when these are expressed in units of their corresponding stellar value. First, normalisation brings different elemental ratios to the same scale, making it easier to compare them and study their mutual relations. As discussed below, such relations provide unique insights into the formation and migration histories of giant planets. Second, normalisation allows us to directly measure the deviation of the planet's composition from that of its host star, which in turn reflects the original composition of the formation environment. As a result, we can constrain the formation history of the planet even in the absence of detailed knowledge of the specific properties of the natal environment. The use of normalised elemental ratios is therefore key to comparing the formation histories of giant planets in multi-planet systems and of giant planets orbiting different stars. 

	\item In all the disc chemical scenarios, the joint evaluation of C- and N-based elemental ratios allows the unequivocal constraining of the source of planetary metallicity. Specifically, gas-dominated giant planets are characterised by N/O*$>$C/O*$>$C/N*, while solid-enriched giant planets are characterised by C/N*$>$C/O*$>$N/O*. Moreover, all the planetary elemental ratios, except C/O*, exhibit monotonic trends with respect to the starting location of the planet. Specifically, the deviation of the ratios from their stellar value increases with the extent of disc-driven migration, independent of the disc chemical scenario and the ionisation level.  
	
	\item The global trends of the elemental ratios in the planet composition through the different migration scenarios are not model-dependent, only their absolute values are so. Specifically, changing the chemical and thermal profiles of the disc only affects the slope of the curves shown in our plots, but not their overall behaviour, as verified by \citet{biazzo2022}. Therefore, our methodology can be directly applied to compare the formation and migration histories of different giant planets in the same planetary system. 
 	
 	\item For solid-enriched giant planets, the C/O* ratio provides constraints only on the accretion scenario, as its value changes limitedly with migration. For gas-dominated ones, it can also provide information also on the chemical structure of the disc, although this is true only for large-scale migration scenarios. 
 
 	\item For gas-dominated giant planets, the joint evaluation of the N/O* and the C/O* ratios allows the constraining of the disc chemical scenario. Specifically, the N/O* ratio is up to a factor of 3 higher in the inheritance scenarios than in the reset ones. In parallel, for large-scale migrations the C/O* ratio is superstellar in the inheritance scenarios and substellar in the reset ones. The elemental ratios of solid-enriched giant planets are instead limitedly affected by whether the disc is characterised by chemical inheritance or reset, making it hard to discriminate between the two scenarios. 
 	
 	\item The S/N* ratio provides additional constraints on the migration history of solid-enriched giant planets. Specifically, large-scale migrations associated with substantial accretion of solids result in S/N*$>$C/N*. On the contrary, short-scale migrations associated with a limited supply of solids and increased contribution of gas accretion result in S/N*$<$C/N*. Such relations holds for all the disc chemical scenarios in which the bulk of N remains in gaseous form while the bulk of S is early trapped in refractory material, so that the planetary S/N* ratio becomes a direct tracer of the planetesimal accretion. 
 	
	\item Given the high volatility contrast between S and N, the joint evaluation of the planetary S/N* ratio and metallicity allows for the distinguishing of hot Jupiters that formed in situ and close to the star from those that underwent extreme migrations, from beyond the snowline of N$_2$. The first will be characterised by stellar values of both the metallicity and the S/N* ratio, while the second will have a markedly superstellar metallicity and stellar S/N* ratio. The metallicity, combined with the C/N* and the S/N* ratios, also allows the tracing of the formation histories of giant planets in discs whose gas phase is enriched in heavy elements due to the sublimation of inwardly drifting dust and pebbles. Such planets will be characterised by superstellar envelope metallicity and C/N*$>$S/N*.

 	\item Pairwise comparison of multiple normalised elemental ratios provides additional insights into the planet formation history and the chemical initial conditions of the disc. The associated binary plots allow us to immediately identify the conditions that favour one scenario over the others, hence providing an effective tool for the interpretation of observational data.
 	
\end{itemize}

\paragraph{Acknowledgements.}
The authors wish to thank Aldo Bonomo, Gloria Guilluy and Ilaria Carleo for their discussion and feedback on exoplanetary atmospheric observations. The authors acknowledge the support of the European Research Council via the Horizon 2020 Framework Programme ERC Synergy “ECOGAL” Project GA-855130, of the Italian National Institute of Astrophysics (INAF) through the INAF Main Stream project ‘Ariel and the astrochemical link between circumstellar discs and planets’ (CUP: C54I19000700005), and of the Italian Space Agency (ASI) through the ASI-INAF contract No. 2021-5-HH.0. The authors also acknowledge the contribution from PRIN INAF 2019 through the project “HOT-ATMOS”. The computational resources for this work were supplied by the Genesis cluster at INAF-IAPS and the technical support of Scigé John Liu is gratefully acknowledged.

\appendix

\section{Planet formation simulations}\label{app:A}

The $N$-body simulations performed in Paper~I investigate the gas and planetesimal accretion histories of giant planets starting their formation tracks at different radial distances from the host star, within their native protoplanetary disc. The initial orbital regions of the giant planets were chosen to explore the possible compositional signatures of the extended planet-forming regions revealed by recent ALMA surveys (see Section~\ref{sec:model} and Paper~I for discussion and references). The $N$-body simulations were performed with the parallel $N$-body code Mercury-Ar$\chi$es \citep{turrini2019,turrini2021}, which accounts for the effects of the disc gas on the dynamical evolution of the planetesimals and implements a two-phase approach to model both the mass growth and the migration of the forming planets. This two-phase approach is modelled after the growth and migration tracks from \citet{bitsch2015}, \citet{dangelo2021}, and \citet{mordasini2015}. The numerical treatment of the physical effects accounted for by Mercury-Ar$\chi$es is summarised here.

The first phase corresponds to the growth of the core and the accretion of its extended atmosphere. During this first phase the giant planet is assumed to undergo migration according to a damped Type~I regime. The second phase corresponds to the runaway gas accretion and the decrease of the planetary radius due to the gas infall. During this second phase the giant planet migrates first by full Type~I migration followed by Type~II migration. The equations governing the evolution of the giant planet during these two phases are described in chronological order the following.

During the first phase of core growth, the planetary mass grows from a Mars-like planetary embryo (M$_{0}$~=~0.1~M$_{\oplus}$) to a critical value of M$_{c}$~=~30~M$_\oplus$, following the growth curve \citep{turrini2011}: 
\begin{equation}
M_1(t)=M_{0}+\left( \frac{e}{e-1}\right)\left(M_{c}-M_{0}\right)\left( 1-e^{-t/\tau_{c}} \right),
\label{eq.mass1}
\end{equation}
where $\tau_{c}$ is the duration of the first growth phase, set to 2~Myr. In parallel, the planetary radius grows following the treatment for extended atmospheres of \citet{fortier2013}, based on the hydrodynamical simulations of \citet{lissauer2009}, as: 
\begin{equation}
R_1(t) = \frac{G\,M_1(t)}{c_{s}^{2}/k_{1}+G\,M_1(t)/\left(k_{2}R_{H}\right)},
%{\frac{c_{s}^{2}}{k_{1}}+\frac{GM_P}{k_{2}R_{H}}}
\label{eqn-inflatedradius}
\end{equation}
where $G$ is the gravitational constant, $c_{s}$ is the speed of sound in the protoplanetary disc at the orbital distance of the planet, $R_{H}$ is the planetary Hill's radius, $k_{1}=1$, and $k_{2}=1/4$ \citep{lissauer2009}. The initial orbit of the giant planet is planar ($i=0^\circ$) and characterised by low eccentricity ($e=10^{-3}$). The damped Type I migration of the giant planet is described by the drift rate \citep{han2005,walsh2011,turrini2021}:
\begin{equation}
\Delta v_{1} = \frac{1}{2}\frac{\Delta a_{1}}{a_{p}}\frac{\Delta t}{\tau_{c}}v_{p},
\label{eqn-typeImigration}
\end{equation}
where $\Delta t$ is the timestep of the $N$-body simulation, $\Delta a_{1}$ is the radial displacement during the growth of the core, and $v_{p}$ and $a_{p}$ are the instantaneous planetary orbital velocity and semimajor axis, respectively. In all $N$-body simulations $\Delta a_{1}$ accounts for 40\% of the total radial displacement \citep{mordasini2015,turrini2021}.

During the second phase of runaway gas accretion, the planetary mass evolves as \citep{turrini2011}:
\begin{equation}
M_{2}(t)=M_{c}+\left( M_{F} - M_{c}\right)\left( 1-e^{-(t-\tau_{c})/\tau_{g}}\right),
\label{eqn-gasgrowth}
\end{equation}
where M$_{F}$ is the final mass of the giant planet set to 317.8~M$_\oplus$, i.e., the Jovian mass, and $\tau_{g}$ is the e-folding time of the runaway gas accretion process. The value of $\tau_{g}$ is set to 0.1~Myr based on the results of hydrodynamical simulations \citep{lissauer2009,dangelo2021}, meaning that the gas giant reaches more than 99\% of its final mass in about 0.5~Myr from the onset of the runaway gas accretion. Once the giant planet enters the runaway gas accretion phase (i.e., for $t > \tau_{c}$), the gravitational infall of the gas causes the planetary radius to shrink and evolve as:
\begin{equation}
R_{2}(t) = R_{E} - \Delta R \left(1-\exp^{-(t-\tau_{c})/\tau_{g}}\right),
\end{equation}
where $R_{E}=R_{1}(\tau_{c}$) is the planetary radius at the end of the core growth phase and $\Delta R = R_{E} - R_{I}$ is the decrease of the planetary radius during the gravitational collapse of the gas. $R_{I}$ is the final inflated radius of the young and hot giant planet assumed to be equal to 1.6~$R_{J}$, where $R_{J}$ is the Jovian radius, based on the hydrodynamical simulations by \citet{lissauer2009} and \citet{dangelo2021}. While in this second formation phase, the migrating giant planet transitions to a full Type~I regime and Type~II regime and its drift rate is \citep{han2005,shibata2020,turrini2021}:
\begin{equation}
\Delta v_{2} = \frac{1}{2}\frac{\Delta a_{2}}{a_{p}}\frac{\Delta t}{\tau_{g}}\exp^{-\left(t-\tau_{c}\right)/\tau_{g}} v_{p},
\label{eqn-typeIImigration}
\end{equation}
where $\Delta a_{2}$ is the radial displacement during the runaway gas accretion. In all $N$-body simulations $\Delta a_{2}$ accounts for 60\% of the total radial displacement \citep{mordasini2015,turrini2021}.

The circumstellar disc considered in the $N$-body simulations of Paper~I is modelled adopting the surface density profile of the protoplanetary disc surrounding the A-type star HD\,163296 \citep{isella2016}, one of the most studied and best characterised discs to date \citep[e.g.,][and references therein]{turrini2021,turrini2021b}. The circumstellar disc has characteristic radius $r_{c}$~=~165 au and gas surface density:
\begin{equation}
\Sigma(r)=\Sigma_{0} \left( r/r_{c} \right)^\gamma exp\left[-\left(r/r_{c} \right)^{\left( 2-\gamma \right)}\right],
\label{eqn-diskdensity}
\end{equation}
where $\gamma=0.8$ \citep{isella2016}. As HD\,163296 is more massive than the Sun, the $\Sigma_{0}$ value was scaled down to 3.3835~g\,cm$^{-3}$ so that the total disc mass matches that of a minimum mass solar nebula \citep{hayashi1981} with the same radial extension, i.e., M$_{*}$~=~0.053 M$_\odot$. The disc gas mass is assumed in steady state and does not decline over time. The inner edge of the disc is set to 0.1~au in all simulations. The disc temperature profile on the midplane is $T(r)= T_{0}\,r^{-0.6}$, where $T_{0}$~=~200~K \citep{andrews2007,oberg2011,eistrup2016}. 

During the runaway gas accretion phase, the giant planet forms a gap in the disc gas whose width is modelled as $W_{gap} = C\cdot R_{H}$ \citep{isella2016,marzari2018}, where the numerical proportionality factor $C=8$ is taken from \citet{isella2016} and \citet{marzari2018}. The gas density $\Sigma_{gap}(r)$ inside the gap evolves over time with respect to the local unperturbed gas density $\Sigma(r)$ as $\Sigma_{gap}(r) = \Sigma(r)\cdot \exp{\left[-\left(t-\tau_{c}\right)/\tau_{g}\right]}$, where $\tau_{c}$ and $\tau_{g}$ are the same as in Equations~\ref{eq.mass1} and \ref{eqn-gasgrowth} \citep{turrini2021}. 

Planetesimals embedded in the disc gas are affected by both the aerodynamic drag of the gas and the disc self-gravity \citep{weidenschilling1977,ward1981,armitage2009}. The gas drag acceleration $F_D$ is expressed by:
\begin{equation}
F_{D} = \frac{3}{8}\frac{C_{D}}{r_{p}}\frac{\rho_{g}}{\rho_{p}}v_{r}^{2},
\label{eqn-gasdrag}
\end{equation}
where $C_{D}$ is the gas drag coefficient, $\rho_{g}$ is the local density of the gas, $\rho_{p}$, and $r_{p}$ are, respectively, the density and radius of the planetesimals, and $v_{r}$ is the relative velocity between the gas and the planetesimals \citep{weidenschilling1977,armitage2009}. The gas drag coefficient $C_D$ is computed following the treatment described by \citet{nagasawa2019} as a function of the Reynolds ($Re$) and Mach ($Ma$) numbers, to account for both subsonic and supersonic regimes of motion of the planetesimals:
\begin{align}
C_{D} = & \left[\left(\frac{24}{Re}+\frac{40}{10+Re}\right)^{-1}+0.23\,Ma\right]^{-2} \nonumber + \, \frac{2\cdot\left(0.8\,k+Ma\right)}{1.6+Ma},  
\end{align}
where $k$ is equal to 0.4 for $Re < 10^{5}$ and to 0.2 for $Re > 10^{5}$ \citep{nagasawa2019}.

The exciting effect of the disc self-gravity is modelled following the approach of \cite{nagasawa2019}, based on the analytical treatment for axisymmetric thin discs from \cite{ward1981}, whose accuracy has been validated by the hydrodynamical study of \citet{fontana2016}. The force due to the disc self-gravity (F$_\text{SG}$) is given by:
\begin{equation}
F_{SG} = 2 \pi\,G\,\Sigma(r) \sum^{\infty}_{n=0} A_{n}\, \frac{\left(1-k\right)\left(4n+1\right)}{\left(2n+2-k\right)\left(2n-1+k\right)},
\label{eqn-selfgravity}
\end{equation}
where $k=0.8$ is the exponent of the power law in Equation~\ref{eqn-diskdensity} and $A_{n}=\left[ (2n)!\,/\,2^{2n}(n!)^{2} \right]^{2}$ \citep{ward1981,marzari2018}. For this value of $k$ the sum on the right hand of Equation~\ref{eqn-selfgravity} converges to the value $-0.754126$ and Equation~\ref{eqn-selfgravity} becomes:
\begin{equation}
F_{SG} = Z \pi G \Sigma(r), 
\end{equation}
where $Z=-1.508252$ (see \citealt{turrini2021} and \citealt{marzari2018} for further discussion). 

Planetesimals are included in the $N$-body simulations as test particles that do not interact with each other nor influence the dynamical evolution of the forming giant planet, i.e., they do not possess gravitational mass. The test particles are dynamically affected by the giant planet, the central star, and the gas in the circumstellar disc. To properly compute the effect of the gas drag, we attribute the test particles inertial masses computed assuming a characteristic radius of 100 km \citep{klahr2016,johansen2017,pirani2019} and density values of 2.4 and 1.0~g\,cm$^{-3}$ depending on whether the planetesimals originated within or beyond the water snowline (see \citealt{turrini2018,turrini2021} for the physical justification of the adopted values and see below for the implications of the choice of the characteristic size of the planetesimals). The test particles are initially on low eccentricity and low inclination orbits ($e \approx i \leq 10^{-2}$), and they uniformly sample the radial extension of the disc between 1~au and the disc radius $r_{c}$, with a spatial density of 2000 particles/au. Particles that encounter the giant planet at a distance closer than the planetary radius at the time of the encounter are considered to be accreted by the planet (see \citealt{turrini2021} and \citealt{podolak2020} for additional discussion). See Section~\ref{sec:model} and \citet{turrini2021} for a discussion of the accreted mass associated to each close encounter. We refer the reader to Figure~\ref{fig:mosaic} and to its animated version in the online journal for a visualisation of the planet formation process modelled by our simulations.

The migration tracks adopted in the $N$-body simulations of Paper~I and described above are not unique, and the migration histories of giant planets can be associated with different migration rates (see also \citealt{pirani2019}). Similarly, planetesimal discs are characterised by continuous size-frequency distributions of the planetesimals \citep[see, e.g.,][and references therein]{krivov2018,turrini2019} rather than a single characteristic size. However, the investigation of \citet{shibata2020} shows that changes in the migration rate and in the size of the planetesimals translate into an enhancement or a reduction of the planetesimal flux on the giant planet. Specifically, faster migration rates and larger planetesimal sizes increase the planetesimal accretion flux, while slower migration rates and smaller planetesimal radii have the opposite effect. Consequently, changes in these two parameters only affect the planetesimal accretion fluxes quantitatively and not qualitatively (see \citealt{shibata2020} for additional discussion). Furthermore, since realistic planetesimal populations are dominated in mass by the high-end tail of their size-frequency distribution \citep[see, e.g.,][for discussion]{krivov2018,turrini2019}, the choice of adopting the planetesimal characteristic radius of 100~km in the simulations of Paper~I instead of larger values means that the planetesimal fluxes computed from these simulations are conservative estimates.

\bibliography{Bibliography.bib}{}
\bibliographystyle{aasjournal}

\end{document}